%% file: main.tex
 \documentclass[camera,letterpaper,nomarginnotes,nonarrowgutter]{jpaper}

\usepackage{tikz}
\usepackage{mathptmx} 
\usepackage{amsmath,amssymb,amsfonts}
\usepackage{comment}
\usepackage{graphics}
\usepackage{fancyhdr}
\usepackage{booktabs}
\usepackage{pifont}
\usepackage[normalem]{ulem}
\usepackage{cite}
\usepackage{hhline}
\usepackage{xcolor}
\usepackage{setspace}
\usepackage{textcomp}
\usepackage{float}
\usepackage{enumitem}
\usepackage{afterpage}
\usepackage{graphicx}
\usepackage[htt]{hyphenat}
\usepackage{makecell}
\usepackage{xspace}
\usepackage{listings}
\usepackage{multirow}
\usepackage{balance}
\usepackage{glossaries} 
\usepackage{siunitx}
\usepackage{duckuments} 
\usepackage{dblfloatfix} 
\usepackage{subcaption}
\usepackage{rotating}
\usepackage[us,12hr]{datetime}
\usepackage[en-GB, useregional=numeric]{datetime2}

\usepackage[compact]{titlesec}
\usepackage[colorinlistoftodos,prependcaption,textsize=small]{todonotes} 
\usepackage{setspace}
\usepackage{dblfloatfix}    
\usepackage{algorithm2e}
\usepackage{soul}
\usepackage{tcolorbox}

\usepackage[bookmarks=true,breaklinks=true,hidelinks]{hyperref}

\newif\ifcameraready
\camerareadytrue

\newif\ifarxiv 
\arxivtrue

\newif\ifrev
\revtrue

\newif\ifpagenumbers
\pagenumberstrue

\newcounter{version}
\ifcameraready
\else
\fi

\title{Spatial Variation-Aware Read Disturbance Defenses:\\Experimental Analysis of Real DRAM Chips\\and Implications on Future Solutions}

\author{
{Abdullah Giray Ya\u{g}l{\i}k\c{c}{\i}}\qquad%
{Yahya Can Tu\u{g}rul}\qquad
{Geraldo F. Oliveira}\\
{{\.I}smail Emir Y{\"u}ksel}\qquad
{Ataberk Olgun}\qquad
{Haocong Luo}\qquad
{Onur Mutlu}\\
ETH Zürich
}

\input{macros}
\input{glossary}

\ifarxiv
    \pagenumberstrue
\else
    \ifcameraready
        \pagenumbersfalse
    \else 
        \pagenumberstrue
        \fancypagestyle{firstpage}
        {
            \fancyhead{}
            \fancyhead[C]{\textcolor{red}{CONFIDENTIAL DRAFT -- DO NOT DISTRIBUTE -- TO APPEAR IN HPCA'24} \\ \textcolor{MidnightBlue}{\emph{Version \theversion{}.1~---~\today, \ampmtime}} }
        }
    \fi
\fi

\makeatletter
\def\bstctlcite{\@ifnextchar[{\@bstctlcite}{\@bstctlcite[@auxout]}}
\def\@bstctlcite[#1]#2{\@bsphack
  \@for\@citeb:=#2\do{%
    \edef\@citeb{\expandafter\@firstofone\@citeb}%
    \if@filesw\immediate\write\csname #1\endcsname{\string\citation{\@citeb}}\fi}%
  \@esphack}
\makeatother

\begin{document}
\bstctlcite{IEEEexample:BSTcontrol}

\maketitle

\newcommand{\hpcaheight}{0mm}
\ifdefined\eaopen
\renewcommand{\hpcaheight}{12mm}
\fi
\ifpagenumbers
    \pagestyle{plain}
\else 
    \pagestyle{empty}
\fi


\input{sections/00_abstract}

\input{sections/01_introduction}

\input{sections/02_background}
\input{sections/03_motivation}
\input{sections/04_methodology}
\input{sections/05_characterization}
\input{sections/06_adaptation}

\input{sections/07_evaluation}
\input{sections/09_relatedwork}
\input{sections/10_conclusion}

\section*{Acknowledgements}
{
We thank the anonymous reviewers of \om{6}{HPCA 2024} for valuable feedback and the SAFARI Research Group members for {constructive} feedback and the stimulating intellectual \agy{4}{environment.} We acknowledge the generous gift funding provided by our industrial partners ({especially} Google, Huawei, Intel, Microsoft\om{4}{)}, which has been instrumental in enabling the decade-long research we have been conducting on read disturbance in DRAM {in particular and memory systems in general.} This work was \om{6}{in part} supported by the Google Security and Privacy Research Award and the Microsoft Swiss Joint Research Center.
}


\balance{}
\bibliographystyle{IEEEtran}
\bibliography{rh-ref}

\ifarxiv
    \onecolumn
    \appendix

    \input{sections/appendix_a_dimms}
\fi

\end{document}

%% file: macros.tex
\def\UrlBreaks{\do\/\do-\/\do.\/\do:}

\expandafter\def\expandafter\UrlBreaks\expandafter{\UrlBreaks
  \do\a\do\b\do\c\do\d\do\e\do\f\do\g\do\h\do\i\do\j
  \do\k\do\l\do\m\do\n\do\o\do\p\do\q\do\r\do\s\do\t
  \do\u\do\v\do\w\do\x\do\y\do\z\do\A\do\B\do\C\do\D
  \do\E\do\F\do\G\do\H\do\I\do\J\do\K\do\L\do\M\do\N
  \do\O\do\P\do\Q\do\R\do\S\do\T\do\U\do\V\do\W\do\X
  \do\Y\do\Z}
  
\definecolor{amber}{rgb}{1.0, 0.49, 0.0}
\definecolor{awesome}{rgb}{1.0, 0.13, 0.32}
\definecolor{dollarbill}{rgb}{0.52,0.73,0.4}
\definecolor{moegi}{rgb}{0.357, 0.537, 0.188}
\definecolor{burgundy}{rgb}{0.5, 0.0, 0.13}
\definecolor{ballblue}{rgb}{0.13, 0.67, 0.8}
\definecolor{ups-truck}{rgb}{0.53, 0.28, 0.21}
\definecolor{airforceblue}{rgb}{0.36, 0.54, 0.66}
\definecolor{cadmiumgreen}{rgb}{0.0, 0.42, 0.24}
\definecolor{darkcyan}{rgb}{0.0, 0.55, 0.55}
\definecolor{caribbeangreen}{rgb}{0.0, 0.8, 0.6}
\definecolor{flamingopink}{rgb}{0.99, 0.56, 0.67}
\definecolor{jazzberryjam}{rgb}{0.65, 0.04, 0.37}
\definecolor{mediumpersianblue}{rgb}{0.0, 0.4, 0.65}
\definecolor{coolblack}{rgb}{0.0, 0.18, 0.39}
\definecolor{bleudefrance}{rgb}{0.19, 0.55, 0.91}
\definecolor{ao}{rgb}{0.0, 0.0, 1.0}
\definecolor{babyblueeyes}{rgb}{0.63, 0.79, 0.95}
\definecolor{darkwarmgray}{rgb}{0.2,0,0}
\definecolor{brightpink}{rgb}{1.0, 0.0, 0.5}

\newcommand{\circled}[1]{{\tikz[baseline=(char.base)]{\node[shape=circle,inner sep=1pt,fill=black, text=white] (char) {\footnotesize \textbf{#1}};}}}

\newcommand{\coloredcircledletter}[2]{\tikz[baseline=(char.base)]{\node[shape=circle,inner sep=1pt,fill=#1, text=white] (char) {\footnotesize \textbf{#2}};}}

\newcommand{\squishlist}{
 \begin{list}{$\circ$}
  { \setlength{\itemsep}{0pt}
     \setlength{\parsep}{0pt}
     \setlength{\topsep}{0pt}
     \setlength{\partopsep}{0pt}
     \setlength{\leftmargin}{1em}
     \setlength{\labelwidth}{1em}
     \setlength{\labelsep}{0.5em} } }

\newcommand{\squishsublist}{
\begin{list}{$\rightarrow$}
 { \setlength{\itemsep}{0pt}
    \setlength{\parsep}{0pt}
    \setlength{\topsep}{-10em}
    \setlength{\partopsep}{-3pt}
    \setlength{\leftmargin}{1em}
    \setlength{\labelwidth}{1em}
    \setlength{\labelsep}{0.5em} } }

\newcommand{\squishend}{
  \end{list}  }

\newcommand{\head}[1]{\noindent\textbf{#1.}} 

\newcounter{obs}
\newcommand\observation[1]{         \refstepcounter{obs}
   \noindent
   \colorbox{gray!20}{\textbf{Observation \theobs.}} \emph{#1}}
   
\newcounter{take}
\newcommand\takebox[1]{\refstepcounter{take}
    \begin{tcolorbox}[colback=white!25!white,colframe=black!65!white, arc=1pt, boxrule=0.5pt, left=2pt,right=2pt,top=0pt,bottom=0pt,  title=\textbf{{Takeaway \thetake.}}]
        \emph{{#1}}
   \end{tcolorbox}
}

\newcommand{\gf}[2]{\ifnum#1=\value{version}\textcolor{red}{#2}\else{#2}\fi}
\newcommand{\agy}[2]{\ifnum#1=\value{version}\textcolor{blue}{#2}\else{#2}\fi}
\newcommand{\yct}[2]{\ifnum#1=\value{version}\textcolor{purple}{#2}\else{#2}\fi}
\newcommand{\om}[2]{\ifnum#1=\value{version}\textcolor{red}{#2}\else#2\fi}
\newcommand{\agytodo}[2]{\ifnum#1=\value{version}\todo[size=\scriptsize, linecolor=orange, bordercolor=orange, backgroundcolor=white]{\textcolor{blue}{TODO:~#2}}\else{}\fi}
\newcommand{\agycomment}[2]{\ifnum#1=\value{version}\todo[size=\scriptsize, linecolor=orange, bordercolor=orange, backgroundcolor=white]{\textcolor{blue}{Giray:~#2}}\else{}\fi}
\newcommand{\omcomment}[2]{\ifnum#1=\value{version}\todo[size=\scriptsize, linecolor=orange, bordercolor=orange, backgroundcolor=white]{\textcolor{red}{Onur:~#2}}\else{}\fi}

\newcommand{\ominlinecomment}[2]{\ifnum#1=\value{version}{\textcolor{red}{\textbf{[Onur:}~#2\textbf{]}}}\else{}\fi}
\newcommand{\versionedparam}[2]{\ifnum#1=\value{version}\textcolor{red}{#2}\else{#2}\fi}
\newcommand{\param}[1]{\versionedparam{0}{#1}}
\newcommand{\N}[0]{\versionedparam{\value{version}}{XX}}
\newcommand{\X}[0]{Svärd}

\newcommand{\secref}[1]{§\ref{#1}}

\newcommand{\tabref}[1]{Table~\ref{#1}}

\newcommand{\figref}[1]{Fig.~\ref{#1}}

\newcommand{\obsref}[1]{Obsv.~\ref{#1}}
\newcommand{\obssref}[1]{Obsvs.~\ref{#1}}

\newcommand{\revtag}[1]{}
\newcommand{\copied}[2]{#2}
\ifcameraready
    \renewcommand{\copied}[2]{#2}
\else
    \ifrev 
        \renewcommand{\revtag}[1]{\todo{\footnotesize #1}}
    \else
        \renewcommand{\copied}[2]{\todo{Copied from #1}\textcolor{gray}{#2}}
    \fi
\fi

\newcommand{\rhmemisolationrefs}[0]{\cite{fournaris2017exploiting, poddebniak2018attacking, tatar2018throwhammer, carre2018openssl, barenghi2018software, zhang2018triggering, bhattacharya2018advanced, google-project-zero, kim2014flipping, rowhammergithub, seaborn2015exploiting, van2016drammer, gruss2016rowhammer, razavi2016flip, pessl2016drama, xiao2016one, bosman2016dedup, bhattacharya2016curious, burleson2016invited, qiao2016new, brasser2017can, jang2017sgx, aga2017good, mutlu2017rowhammer, tatar2018defeating, gruss2018another, lipp2018nethammer, van2018guardion, frigo2018grand, cojocar2019eccploit,  ji2019pinpoint, mutlu2019rowhammer, hong2019terminal, kwong2020rambleed, frigo2020trrespass, cojocar2020rowhammer, weissman2020jackhammer, zhang2020pthammer, yao2020deephammer, deridder2021smash, hassan2021utrr, jattke2022blacksmith, tol2022toward, kogler2022half, orosa2022spyhammer, zhang2022implicit, liu2022generating, cohen2022hammerscope, zheng2022trojvit, fahr2022frodo, tobah2022spechammer, rakin2022deepsteal, park2016statistical, park2016experiments,lim2017active, ryu2017overcoming, yun2018study, yang2019trap, walker2021ondramrowhammer, kim2020revisiting, orosa2021deeper, yaglikci2022understanding, khan2018analysis, agarwal2018rowhammer, li2014write, ni2018write, genssler2022reliability, mutlu2023fundamentally}}

\newcommand{\exploitingRowHammerAllCitations}[0]{\cite{fournaris2017exploiting, poddebniak2018attacking, tatar2018throwhammer, carre2018openssl, barenghi2018software, zhang2018triggering, bhattacharya2018advanced, google-project-zero, kim2014flipping, rowhammergithub, seaborn2015exploiting, van2016drammer, gruss2016rowhammer, razavi2016flip, pessl2016drama, xiao2016one, bosman2016dedup, bhattacharya2016curious, burleson2016invited, qiao2016new, brasser2017can, jang2017sgx, aga2017good, mutlu2017rowhammer, tatar2018defeating, gruss2018another, lipp2018nethammer, van2018guardion, frigo2018grand, cojocar2019eccploit,  ji2019pinpoint, mutlu2019rowhammer, hong2019terminal, kwong2020rambleed, frigo2020trrespass, cojocar2020rowhammer, weissman2020jackhammer, zhang2020pthammer, yao2020deephammer, deridder2021smash, hassan2021utrr, jattke2022blacksmith, tol2022toward, kogler2022half, orosa2022spyhammer, zhang2022implicit, liu2022generating, cohen2022hammerscope, zheng2022trojvit, fahr2022frodo, tobah2022spechammer, rakin2022deepsteal, aydin2022cyber, mus2022jolt, wang2022research, lefforge2023reverse,fahr2022effects, kaur2022work, cai2022feasibility, li2022cyberradar, roohi2022efficient, staudigl2022neurohammer, yang2022socially, islam2022signature, tomita2022extracting, france2022modeling}}

\newcommand{\understandingRowHammerAllCitations}[0]{\cite{redeker2002investigation, kim2014flipping, park2014active, park2016statistical, yang2016suppression, park2016experiments,lim2017active, ryu2017overcoming, yang2017scanning, lim2018study, yun2018study, yang2019trap, gautam2019row, walker2021ondramrowhammer, kim2020revisiting, orosa2021deeper, jiang2021quantifying, orosa2022spyhammer, cohen2022hammerscope, yaglikci2022understanding, khan2018analysis, agarwal2018rowhammer, li2014write, ni2018write, genssler2022reliability, mutlu2023fundamentally, he2023whistleblower, baeg2022estimation, frigo2020trrespass, mutlu2017rowhammer, mutlu2018rowhammer, mutlu2019rowhammer, olgun2023hbm, olgun2023drambender, zhou2023double, luo2023rowpress}}

\newcommand{\mitigatingRowHammerAllCitations}[0]{\cite{AppleRefInc, rh-hp,rh-lenovo,greenfield2012throttling, kim2014flipping, kim2014architectural, bains14d, bains14c, jedec2017ddr4, aichinger2015ddr, aweke2016anvil, bains-merged, bains2015row, bains2016distributed, bains2016row, gomez2016dummy, yang2016suppression, son2017making, seyedzadeh2018cbt, irazoqui2016mascat, ryu2017overcoming, yang2017scanning, you2019mrloc, lee2019twice, park2020graphene, yaglikci2021security, yaglikci2021blockhammer, frigo2020trrespass, kang2020cattwo, hassan2021utrr, qureshi2022hydra, saileshwar2022randomized, brasser2017can, konoth2018zebram, van2018guardion, vig2018rapid, hassan2019crow, gautam2019row, kim2022mithril, lee2021cryoguard, marazzi2022protrr, zhang2022softtrr, joardar2022learning, juffinger2023csi, yaglikci2022hira, saxena2022aqua, manzhosov2022revisiting, ajorpaz2022evax, naseredini2022alarm, joardar2022machine, hassan2022case, zhang2020leveraging,loughlin2021stop, devaux2021method, han2021surround, fakhrzadehgan2022safeguard, saroiu2022price, saroiu2022configure, loughlin2022moesiprime, zhou2022lt, hong2023dsac, mutlu2023fundamentally, marazzi2023rega, di2023copy, sharma2022review, woo2023scalable, park2022row, wi2023shadow, kim2023ddr5, gude2023defending, guha2022criticality, france2022modeling, france2022reducing, bennett2021panopticon, enomoto2022efficient, arikan2022processor, tomita2022extracting, saxena2023pt, zhou2023dnndefender, bostanci2024comet, olgun2024abacus}}

\newcommand{\hwBasedRowHammerMitigations}[0]{\cite{AppleRefInc, rh-hp,rh-lenovo,greenfield2012throttling, kim2014flipping, kim2014architectural, bains14d, bains14c, aichinger2015ddr, aweke2016anvil, bains-merged, bains2015row, bains2016distributed, bains2016row, son2017making, irazoqui2016mascat, ryu2017overcoming, yang2017scanning, seyedzadeh2017cbt, you2019mrloc, lee2019twice, park2020graphene, yaglikci2021blockhammer, frigo2020trrespass, kang2020cattwo, hassan2021utrr, qureshi2022hydra, saileshwar2022randomized, brasser2017can, konoth2018zebram, van2018guardion, vig2018rapid, gautam2019row, kim2022mithril, lee2021cryoguard, zhang2022softtrr, joardar2022learning, juffinger2023csi, yaglikci2022hira, saxena2022aqua, enomoto2022efficient, manzhosov2022revisiting, ajorpaz2022evax, joardar2022machine, hassan2022case, zhang2020leveraging, loughlin2021stop, devaux2021method, han2021surround, fakhrzadehgan2022safeguard, saroiu2022price, saroiu2022configure, loughlin2022moesiprime, zhou2022lt, mutlu2023fundamentally, di2023copy, sharma2022review, woo2023scalable, park2022row, wi2023shadow, kim2023ddr5, gude2023defending, guha2022criticality, france2022modeling, france2022reducing, arikan2022processor, tomita2022extracting, saxena2023pt, zhou2023dnndefender, yang2016suppression, bostanci2024comet, bennett2021panopticon, seyedzadeh2018cbt, gomez2016dummy, gautam2019rowhammering, wang2021discreet, hassan2019crow, yaglikci2021security, woo2022scalable, zhou2022ltpim, olgun2024abacus, marazzi2022protrr}}

\newcommand{\swBasedRowHammerMitigations}[0]{
\cite{enomoto2022efficient, konoth2018zebram, van2018guardion, brasser2017can, bock2019riprh, aweke2016anvil, zhang2022softtrr}
}

\newcommand{\mcBasedRowHammerMitigations}[0]{\cite{AppleRefInc, rh-hp,rh-lenovo,greenfield2012throttling, kim2014flipping, kim2014architectural, bains14d, bains14c, aichinger2015ddr, aweke2016anvil, bains-merged, bains2015row, bains2016distributed, bains2016row, son2017making, irazoqui2016mascat, ryu2017overcoming, yang2017scanning, seyedzadeh2017cbt, you2019mrloc, lee2019twice, park2020graphene, yaglikci2021blockhammer, frigo2020trrespass, kang2020cattwo, hassan2021utrr, qureshi2022hydra, saileshwar2022randomized, brasser2017can, konoth2018zebram, van2018guardion, vig2018rapid, gautam2019row, kim2022mithril, lee2021cryoguard, zhang2022softtrr, joardar2022learning, juffinger2023csi, yaglikci2022hira, saxena2022aqua, enomoto2022efficient, manzhosov2022revisiting, ajorpaz2022evax, joardar2022machine, hassan2022case, zhang2020leveraging, loughlin2021stop, devaux2021method, han2021surround, fakhrzadehgan2022safeguard, saroiu2022price, saroiu2022configure, loughlin2022moesiprime, zhou2022lt, mutlu2023fundamentally, di2023copy, sharma2022review, woo2023scalable, park2022row, wi2023shadow, kim2023ddr5, gude2023defending, guha2022criticality, france2022modeling, france2022reducing, arikan2022processor, saxena2023pt, zhou2023dnndefender}}

\newcommand{\inDRAMRowHammerMitigations}[0]{\cite{jedec2017ddr4, gomez2016dummy, seyedzadeh2018cbt, yang2016suppression, yaglikci2021security, marazzi2022protrr, hassan2019crow, hong2023dsac, marazzi2023rega, bennett2021panopticon}}

\newcommand{\refreshBasedRowHammerDefenseCitations}[0]{\cite{rh-lenovo, rh-hp, bains14c, bains14d, bains-merged, lee2019twice, seyedzadeh2017cbt,vig2018rapid, irazoqui2016mascat, seyedzadeh2018cbt, kang2020cattwo, park2020graphene, kim2022mithril, kim2014architectural, bains2015row, bains2016distributed, bains2016row, aweke2016anvil, AppleRefInc, kim2014flipping, son2017making, you2019mrloc, yaglikci2021security, frigo2020trrespass, hassan2021utrr, loughlin2021stop, qureshi2022hydra, devaux2021method, wang2021discreet, marazzi2022protrr, zhang2022softtrr, joardar2022learning, yaglikci2022hira, saroiu2022configure, bostanci2024comet, olgun2024abacus, joardar2022machine}}

\newcommand{\throttlingBasedRowHammerDefenseCitations}[0]{\cite{greenfield2012throttling, yaglikci2021blockhammer}}

\newcommand{\isolationBasedRowHammerDefenseCitations}[0]{\cite{saileshwar2022randomized, wi2023shadow, loughlin2021stop, zhou2023dnndefender, woo2023scalable, bock2019riprh,konoth2018zebram, van2018guardion, brasser2017can, saxena2022aqua, woo2022scalable, hassan2019crow}}

\newcommand{\circuitBasedRowHammerDefenseCitations}[0]{\cite{park2022row, kim2023ddr5, gautam2019row, yang2016suppression, hassan2019crow, gomez2016dummy, han2021surround, ryu2017overcoming, yang2017scanning, zhou2022ltpim, lee2021cryoguard, hassan2022case, gautam2019rowhammering}}

\newcommand{\rowHammerGetsWorseCitations}[0]{\cite{kim2014flipping, kim2020revisiting, frigo2020trrespass, yaglikci2022understanding, orosa2021deeper, mutlu2017rowhammer, mutlu2018rowhammer, mutlu2019rowhammer, cojocar2020rowhammer, mutlu2023fundamentally, luo2023rowpress}}

\newcommand{\rowHammerDefenseScalingProblemsCitations}[0]{\cite{kim2020revisiting, yaglikci2021blockhammer, park2020graphene, mutlu2017rowhammer, mutlu2018rowhammer, mutlu2019rowhammer, mutlu2023fundamentally, hassan2021utrr}}



%% file: glossary.tex
\newcommand{\hcfirst}[0]{HC_{first}}
\newacronym{hcfirst}{$\hcfirst$}{the minimum {hammer count} {required to induce the first bitflip}} 
\newacronym{ber}{$BER$}{bit error rate}
\newacronym{wcdp}{$WCDP$}{worst-case data pattern}
\newcommand{\nummodules}{\param{15}}
\newcommand{\numchips}{\agy{2}{144}}
\newcommand{\numworkloadmixes}{\agy{4}{120}}
\newcommand{\wlcnt}{\numworkloadmixes{}}
\newacronym{taggon}{$t_{AggOn}$}{the time that an aggressor row stays active}
\newacronym{taggoff}{$t_{AggOff}$}{the time that an aggressor row stays precharged}
\newacronym{tras}{$t_{RAS}$}{charge restoration latency}
\newacronym{trp}{$t_{RP}$}{precharge latency}
\newcommand{\trc}[0]{t_{RC}}
\newacronym{trc}{$\trc{}$}{row activation cycle}
\newacronym{trcd}{$t_{RCD}$}{row activation latency}
\newacronym{tcl}{$t_{CL}$}{column access latency}
\newacronym{tcwl}{$t_{CWL}$}{column write latency}
\newcommand{\tfaw}[0]{t_{FAW}}
\newacronym{tfaw}{$\tfaw{}$}{four row activation window}
\newcommand{\trefw}[0]{t_{REFW}}
\newacronym{trefw}{$\trefw{}$}{refresh window}
\newcommand{\trefi}[0]{t_{REFI}}
\newacronym{trefi}{$\trefi{}$}{refresh interval}
\newacronym{trrslack}{$t_{RefSlack}$}{{the maximum delay between the time a {periodic}/{preventive} refresh is generated and the time the refresh is performed}}
\newacronym{tapa}{$t_{APA}$}{the latency of issuing $ACT-PRE-ACT$ command sequence}
\newacronym{ref}{$REF$}{refresh}
\newacronym{act}{$ACT$}{activate}
\newacronym{pre}{$PRE$}{precharge}
\newcommand{\trfc}[0]{t_{RFC}}
\newacronym{trfc}{$\trfc{}$}{refresh latency}
\newacronym{iqr}{$IQR$}{interquartile range}
\newacronym{cv}{$CV$}{the coefficient of variation}
\newacronym{hc}{$HC$}{hammer count}
\newcommand{\pth}[0]{p_{th}}
\newacronym{pth}{$\pth{}$}{{PARA's probability threshold}}
\newcommand{\pf}[0]{p_{failure}}
\newacronym{pf}{$\pf{}$}{failure probability over a sufficiently long time}
\newcommand{\prh}[0]{p_{RH}}
\newacronym{prh}{$\prh{}$}{reliability target for a \gls{trefw}}

\newcommand{\cchip}[0]{D_{chip}}
\newacronym{cchip}{$\cchip{}$}{chip density}

\newcommand{\rbcpki}[0]{RBCPKI}
\newacronym{rbcpki}{$\rbcpki{}$}{row buffer conflicts per kilo instruction}

\newcommand{\mpki}[0]{MPKI}
\newacronym{mpki}{$\mpki{}$}{misses per kilo instruction}

\newcommand{\vdd}[0]{V_{DD}}
\newacronym{vdd}{$\vdd{}$}{supply voltage}

\newcommand{\gnd}[0]{GND}
\newacronym{gnd}{$\gnd{}$}{ground}

\newcommand{\rd}[0]{RD}
\newacronym{rd}{$\rd{}$}{read}

\newcommand{\dramwr}[0]{WR}
\newacronym{wr}{$\dramwr{}$}{write}

%% file: sections/00_abstract.tex
\begin{abstract}
Read disturbance in modern DRAM chips is a widespread \om{2}{phenomenon} and is reliably used for breaking memory isolation, \om{2}{a} fundamental building block \om{2}{for building robust systems}. RowHammer and RowPress are two examples of read disturbance in DRAM where repeatedly accessing (hammering) or keeping active (pressing) a memory location induces bitflips in other memory locations. Unfortunately, shrinking technology node size exacerbates read disturbance in DRAM chips over generations. As a result, existing defense mechanisms suffer from significant performance \om{2}{and} energy overheads, limited effectiveness, or prohibitively high hardware complexity. 
    
In this paper, we tackle these shortcomings by leveraging the spatial variation in read disturbance across different memory locations in real DRAM chips.
To do so, we
1)~present the first rigorous real DRAM chip characterization study of spatial variation of read disturbance and
2)~propose Svärd, a new mechanism that dynamically adapts the aggressiveness of existing solutions based on the \agy{2}{row-level} read disturbance profile. 
Our experimental characterization on \numchips{} real DDR4 DRAM chips representing 10 \agy{7}{chip designs} demonstrates a large variation in read disturbance vulnerability across different memory locations\om{4}{:} \agy{4}{in the part of memory with the worst read disturbance vulnerability, 1)~up to $2\times$ the number of bitflips can occur and 2)~bitflips can occur at an order of magnitude fewer accesses, compared to the memory locations with the least vulnerability to read disturbance.} 
Svärd \om{2}{leverages \om{4}{this} variation to reduce} the overheads of \param{five} state-of-the-art read disturbance \agy{4}{solutions}, \om{2}{and thus significantly increases system performance.} 
\end{abstract}

%% file: sections/01_introduction.tex
\section{Introduction}
\label{sec:introduction}

{To ensure system \om{2}{robustness (including} reliability, security, and safety\om{2}{)}, it is critical to {maintain} {memory isolation: accessing a memory address should not cause unintended side-effects on data stored on other addresses~\cite{kim2014flipping}. {Unfortunately}, with aggressive technology scaling, DRAM~\cite{dennard1968dram}, the prevalent {main} memory technology}, suffers from increased \emph{read disturbance}: accessing (reading) a row of DRAM cells (\om{2}{i.e., a} DRAM row) degrades the data integrity of other physically close but \emph{unaccessed} DRAM rows.}
\emph{RowHammer} and \emph{RowPress} are two prime examples of \om{2}{the} {DRAM read {disturbance} phenomenon {where} a DRAM row (i.e., victim row) can experience bitflips when a nearby DRAM row (i.e., aggressor row) is} 1)~repeatedly opened (i.e., hammered)~\rhmemisolationrefs{} or 2)~kept open for a long period (i.e., pressed)~\cite{luo2023rowpress}, respectively.

Many prior works demonstrate attacks on a wide range of systems \om{4}{that} exploit read disturbance {to escalate {privilege}, leak private data, and manipulate critical application outputs~\exploitingRowHammerAllCitations{}.
To make matters worse, \om{2}{various} experimental studies~\cite{kim2014flipping, mutlu2017rowhammer, mutlu2019rowhammer, frigo2020trrespass, cojocar2020rowhammer, kim2020revisiting, kim2014flipping, luo2023rowpress} \om{2}{find} that {newer DRAM chip generations are more susceptible to read disturbance}. {For example,} chips manufactured in {2018-2020} can experience RowHammer {bitflips} at an order of magnitude fewer row activations compared to the chips manufactured in {2012-2013}~\cite{kim2020revisiting}. As {read disturbance \agy{2}{in} DRAM chips} worsens, ensuring \om{2}{robust (i.e.,} \agy{2}{reliable, secure, and safe}\om{2}{)} operation {becomes} more expensive in terms of performance overhead, energy consumption, and hardware complexity~\cite{kim2020revisiting, yaglikci2021blockhammer, park2020graphene}. Therefore, it is {critical} to understand {the read disturbance \om{4}{vulnerabilities of} DRAM chips} in greater detail with in-depth insights \om{2}{in order}
to develop more effective and {efficient} solutions for current and future DRAM-based {memory} systems.}
To this end, prior works study \om{2}{various} characteristics of DRAM read disturbance~\understandingRowHammerAllCitations{}. Unfortunately, \emph{no} prior work rigorously studies the spatial variation {of} DRAM read disturbance and its implications on future solutions.

Our \emph{goal} in this paper is to build a detailed understanding of the spatial variation in read disturbance across DRAM rows and leverage this understanding to improve the existing solutions.
To this end, in this paper, we 1)~present the first rigorous experimental characterization of the spatial variation in read disturbance \agy{4}{vulnerabilities of} \om{2}{real} DRAM \om{2}{chips}, and 2)~propose \X{}, a new mechanism that dynamically adapts the aggressiveness of existing solutions based on the \om{2}{observed} variation.

\vspace{-0.2em}
\head{Characterization} We test \numchips{} modern real DDR4 DRAM chips representing \om{4}{\param{10}} different \agy{7}{chip designs} from three major manufacturers.\footnote{We test DRAM chips from Samsung, SK Hynix, and Micron, which hold the largest market shares of  40.7\%, 28.8\%, and 26.4\% respectively~\cite{alsop2023DRAM, wang2022Falling}. \agy{7}{Chip design is identified by chip density, die revision, and chip organization.}} Our experimental results show that the \agy{4}{read disturbance vulnerabilities of DRAM chips vary} \emph{significantly} and \emph{irregularly} across 1)~DRAM rows within a DRAM subarray, i.e., a two-dimensional DRAM array that consists of hundreds of DRAM rows~(\secref{sec:background}) and 2)~subarrays within a DRAM bank, i.e., a group of many subarrays that share an I/O circuitry~(\secref{sec:background}).
\agy{2}{We \agy{4}{make two key observations from our characterization: 1)~the} fraction of erroneous DRAM cells, i.e., \gls{ber}, \om{4}{varies by a factor of $2\times$ and 2)~\gls{hcfirst} varies} by an order of magnitude across DRAM rows in a subarray.} 
\vspace{-0.2em}

\agy{4}{We} investigate \om{4}{the} correlations between a victim DRAM row's spatial features (i.e., physical location within the corresponding subarray and bank) and read disturbance vulnerability \agy{4}{in two steps. First,} we reverse engineer the organization of a DRAM bank. 
\agy{4}{Second, we} perform \om{4}{extensive} statistical analyses based on \agy{4}{1)}~reverse-engineered spatial features and \agy{4}{2)}~observed \gls{hcfirst} values. \agy{4}{We make the key observation} that the variation in read disturbance vulnerability across DRAM rows correlates well with the spatial features of DRAM rows \emph{only} in four out of \param{15} tested DRAM modules. \agy{4}{Therefore, we conclude that read disturbance vulnerability irregularly varies across DRAM rows, and thus, row-level spatial features \om{5}{we evaluate} are insufficient to predict read disturbance vulnerability.}

\head{Improving existing solutions} 
\om{4}{Based on the understanding we develop via our rigorous experimental characterization,} we propose \X{}\om{2}{,} a new mechanism that dynamically adapts the aggressiveness of existing solutions \agy{2}{to the read disturbance vulnerability level of \om{5}{each} potential victim row. \X{} forces an existing solution to react more aggressively to an aggressor row activation (e.g., preventively refresh the potential victim row) if \om{4}{a} potential victim row is more vulnerable to read disturbance.} \X{} is implemented together with existing solutions, which can be either 1)~in the memory controller within the processor chip \om{4}{(i.e.,} without \om{2}{modifications to} the DRAM chip \om{2}{or \om{4}{DRAM} interface}\om{4}{)} or 2)~within the DRAM chip \om{4}{(i.e.,} \om{4}{transparent} to the rest of the system\om{4}{)}.
\agy{4}{We evaluate the performance benefits of \X{} on \param{five} state-of-the-art solutions,}
\agy{5}{AQUA~\cite{saxena2022aqua},} {BlockHammer~\cite{yaglikci2021blockhammer},} Hydra~\cite{qureshi2022hydra}, PARA~\cite{kim2014flipping}, and RRS~\cite{saileshwar2022randomized}
\agy{4}{using three representative spatial distribution profiles of read disturbance that we generate based on our experimental characterization \om{5}{of real DRAM chips} (i.e., one for each manufacturer). Our performance evaluation results show that}
\X{} significantly \om{2}{reduces the performance overheads} of 
{\agy{5}{AQUA}, BlockHammer,} Hydra, PARA, and RRS, \om{4}{leading to system performance improvements of} \agy{5}{$1.23\times$,} {$2.65\times$}, $1.03\times$, $1.57\times$, and $2.76\times$, \agy{2}{respectively,} on average across \wlcnt{} multiprogrammed memory-intensive workloads.

{We make the following contributions:}
\begin{itemize}
    \item We present the first rigorous characterization of the spatial variation \om{4}{of} read disturbance in \om{2}{modern} DRAM \om{2}{chips}. Our experimental results on \numchips{} real \om{2}{DDR4} DRAM chips {spanning 10 different \agy{7}{chip designs}} show a \agy{2}{significant and irregular} variation \om{2}{in} read {disturbance} vulnerability across DRAM rows. 
    \item We propose \X{}, a new mechanism that dynamically adapts the aggressiveness of an existing read disturbance solution \agy{2}{to the vulnerability level of the potential victim row.}
    \item We showcase \X{}'s integration with \param{five} different state-of-the-art read disturbance \agy{4}{solutions}. \om{2}{Our results} show that \X{} reduces the performance overhead of \om{2}{these four state-of-the-art} solutions, \om{4}{leading to large system performance benefits.} 
\end{itemize}

%% file: sections/02_background.tex
\section{Background}
\label{sec:background}
This section provides a concise overview of 1)~DRAM organization and operation
and 2)~DRAM read disturbance.
For more detail, we refer the reader to prior works {on DRAM and read disturbance}~\cite{ipek2008self,zhang2014half, qureshi2015avatar, liu2012raidr, liu2013experimental, keeth2001dram, mutlu2007stall, moscibroda2007memory, mutlu2008parbs, kim2010atlas, subramanian2014bliss, salp, kim2014flipping,
hassan2016chargecache, chang2016understanding, lee2017design,  chang2017understanding,  patel2017reaper,kim2018dram, kim2020revisiting, hassan2019crow, frigo2020trrespass, chang2014improving, chang2016low, ghose2018vampire, hassan2017softmc, khan2016parbor, khan2016case, khan2014efficacy, seshadri2015gather, seshadri2017ambit, kim2018solar, kim2019d, patel2019understanding, patel2020beer, lee2013tiered, lee2015decoupled, seshadri2013rowclone, luo2020clrdram, seshadri2019dram, wang2020figaro}.
\agycomment{4}{Non-SAFARI references: \cite{keeth2001dram, ipek2008self,zhang2014half, qureshi2015avatar}}


\subsection{DRAM {Organization and Operation}}
\label{sec:dram_background}

\begin{figure}[h]
    \centering
    \includegraphics[width=\linewidth]{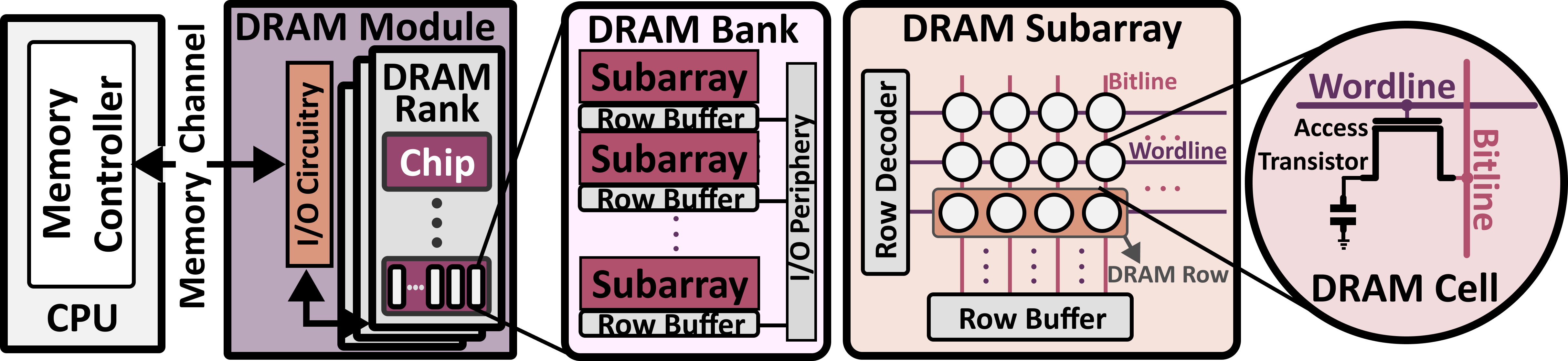}
    \caption{DRAM organization}
    \label{fig:dram_organization}
\end{figure}

\head{Organization}
{\figref{fig:dram_organization}a shows the organization of DRAM-based memory systems. A memory channel connects the processor (CPU) to a set of DRAM chips, called \agy{4}{\emph{DRAM rank}}.}
\agy{4}{Chips in a DRAM rank operate in lock-step.}
\agy{0}{Each chip has multiple DRAM banks, each consisting of multiple DRAM cell arrays \om{4}{(called \emph{subarrays})} and their local I/O circuitry. \om{4}{Within a subarray,} DRAM cells are organized as a two-dimensional array \om{4}{of} DRAM rows and columns.}
{A} DRAM cell {stores one bit of data} {in the form of} electrical charge in {a} capacitor, which can be accessed through an access transistor.  
A wire called \om{4}{wordline} drives the gate of all DRAM cells' access transistors in a DRAM row\om{4}{. A} wire called \om{4}{\emph{bitline}} connects all DRAM cells in a DRAM column to a common differential sense amplifier. Therefore, when a wordline is asserted, each DRAM cell in the DRAM row is connected to its corresponding sense amplifier. The set of sense amplifiers in a subarray is called \om{4}{\emph{the row buffer}}, where the data of \om{4}{an activated} DRAM row is buffered \agy{4}{to serve a column access.}

\head{Operation}
\copied{ABACUS}{
The memory controller serves memory access requests by issuing DRAM commands, e.g., row activation ($ACT$), bank precharge ($PRE$), data read ($RD$), data write ($WR$), and refresh ($REF$) \gf{0}{while respecting certain timing parameters to guarantee correct operation~\cite{jedec2020ddr5,jedec2020lpddr5,jedec2015lpddr4, jedec2015hbm,jedecddr,jedec2017ddr4,jedec2012ddr3}.}
To read or write data\om{4}{;} the memory controller first \om{4}{needs to activate the corresponding row. To do so, it} issues an $ACT$ command alongside the bank address and row address corresponding to the memory request's address. 
When a row is activated, its data is copied \gf{0}{to and temporarily stored at} the row buffer.
\gf{0}{The latency from the start of a row activation until the data
is reliably readable\om{4}{/writable} in the row buffer is called \om{4}{the} \emph{\gls{trcd}}.}
\gf{0}{During the row activation process, a DRAM cell loses its charge, and thus, its initial charge needs to be restored (\om{4}{via} a process called \emph{charge restoration}). 
The latency from the start of a row activation until the completion of the DRAM cell's charge restoration is called \agy{4}{the \emph{\gls{tras}}}.
}
The memory controller can read/{write} data from/to the row buffer using $RD$/$WR$ commands.
The changes are propagated to the DRAM cells in the open row. 
Subsequent accesses to the same row can be served quickly \om{4}{from the row buffer (i.e., called a \emph{row hit})} without issuing another $ACT$ to the same row. 
\gf{0}{The latency of performing a read/write operation is called \gls{tcl}/\gls{tcwl}}. 
To access another row in \om{4}{an already} activated DRAM bank, the memory controller must issue a $PRE$ command \om{4}{to} close the opened row \om{4}{and prepare the bank for a new activation}.
\agy{4}{When the $PRE$ command is issued}, the DRAM chip de-asserts the active row's wordline and precharges the bitlines. The timing parameter for precharge is called \om{4}{the \emph{\gls{trp}}}.}

{\om{4}{A} DRAM cell \om{4}{is} inherently leaky and thus \om{4}{loses its} stored electrical charge over time. To maintain data integrity, a DRAM cell {is periodically refreshed} with a {time interval called \agy{4}{the \emph{\gls{trefw}}}, which is typically} \SI{64}{\milli\second} (e.g.,~\cite{jedec2012ddr3, jedec2017ddr4, micron2014ddr4}) or \SI{32}{\milli\second} (e.g.,~\cite{jedec2015lpddr4, jedec2020ddr5, jedec2020lpddr5}) \agy{4}{at normal operating temperature (\om{5}{i.e.}, up to \SI{85}{\celsius}) and half of it for the extended temperature range (\om{5}{i.e.}, above \SI{85}{\celsius} up to \SI{95}{\celsius})}.  
To {timely} refresh all cells, the memory controller {periodically} issues a refresh {($REF$)} command with {a time interval called} \agy{4}{the \emph{\gls{trefi}}}, {which is typically} \SI{7.8}{\micro\second} {(e.g.,~\cite{jedec2012ddr3, jedec2017ddr4, micron2014ddr4}) or} \SI{3.9}{\micro\second} (e.g.,~\cite{jedec2015lpddr4, jedec2020ddr5, jedec2020lpddr5}) \agy{4}{at normal operating temperature.} When a rank-/bank-level refresh command is issued, the DRAM chip internally refreshes several DRAM rows, during which the whole rank/bank is busy. This operation's latency is called \agy{4}{the \emph{\gls{trfc}}}.} 


\subsection{Read Disturbance in DRAM}
\label{sec:background_rowhammer}
Read disturbance is the phenomenon that reading data from a memory or storage device causes physical disturbance (e.g., voltage deviation, electron injection, electron trapping) on another piece of data that is \emph{not} accessed but physically located nearby the accessed data. Two prime examples of read disturbance in modern DRAM chips are RowHammer~\cite{kim2014flipping}, and RowPress~\cite{luo2023rowpress}, where repeatedly accessing (hammering) or keeping active (pressing) a DRAM row induces bitflips in physically nearby DRAM rows, respectively. In RowHammer and RowPress terminology, the row that is hammered or pressed is called the \emph{aggressor} row, and the row that experiences bitflips the \emph{victim} row.
\copied{ABACUS}{For read disturbance bitflips to occur, 1)~the aggressor row needs to be activated more than a certain threshold value, \agy{0}{defined as \gls{hcfirst}~\cite{kim2020revisiting} \om{4}{and/}or 2)~\gls{taggon}~\cite{luo2023rowpress} need to be large-enough~\cite{kim2020revisiting, orosa2021deeper, yaglikci2022understanding, luo2023rowpress}. To avoid read disturbance bitflips, systems take preventive actions, e.g., \om{4}{they refresh} victim rows~\refreshBasedRowHammerDefenseCitations{}, selectively \om{4}{throttle} accesses to aggressor rows~\cite{yaglikci2021blockhammer, greenfield2012throttling}, and physically \om{4}{isolate} potential aggressor and victim rows~\cite{hassan2019crow, konoth2018zebram, saileshwar2022randomized, saxena2022aqua, wi2023shadow, woo2023scalable}. These \agy{4}{solutions} aim to perform preventive actions before the cumulative effect of an aggressor row's \emph{activation count} and \emph{on time} causes read disturbance bitflips.
}}

%% file: sections/03_motivation.tex
\section{Motivation and Goal}
\label{sec:motivation}
\copied{DeeperLook}{Prior research experimentally demonstrates that read disturbance is {clearly a} worsening DRAM \om{5}{robustness (i.e.,} reliability, security, \om{5}{and safety)} \agy{5}{concern}~\rowHammerGetsWorseCitations{}.
Despite all efforts, {newer} DRAM chips are shown to be \emph{significantly} more vulnerable to read disturbance than older generations~\cite{kim2020revisiting}. Even DRAM chips that have been marketed as RowHammer-free in 2018-2020 experience RowHammer \gf{0}{bitflips} at \emph{significantly} lower \agy{5}{hammer} counts (e.g., {4.8K \om{4}{activations} for each of two aggressor rows} for LPDDR4 chips when \gf{0}{target row refresh (TRR)} protection\footnote{\gf{0}{DRAM manufacturers implement RowHammer \agy{4}{solutions}, generally called Target Row Refresh (TRR)~\cite{frigo2020trrespass,jedec2015hbm,jedec2017ddr4}, which perform proprietary operations within DRAM to prevent RowHammer bitflips}.} is disabled~\cite{kim2020revisiting} and {{2}5K} {for DDR4 chips} when {TRR} protection is enabled~\cite{frigo2020trrespass}) compared to the {DDR3} DRAM chips manufactured in 2012-2013 (e.g., {139K}~\cite{kim2014flipping} \om{4}{or 69K~\cite{kim2020revisiting}}). Many prior works~\mitigatingRowHammerAllCitations{} propose RowHammer \agy{4}{solutions} to provide RowHammer-safe operation with either probabilistic or deterministic security guarantees. Unfortunately, {recent} works~\rowHammerDefenseScalingProblemsCitations{} demonstrate that many of these \agy{4}{solutions} will 
{incur} \emph{significant} performance, energy consumption, and hardware complexity overheads such that they become prohibitively expensive when deployed in future DRAM chips \om{4}{with much larger read disturbance vulnerabilities}~\cite{kim2020revisiting}.}

\copied{DeeperLook}{To avoid read disturbance bitflips in future DRAM-based computing systems in an effective {and} efficient way, it is {critical} to {rigorously} gain {detailed} insights into the read disturbance phenomena 
\agy{4}{under various circumstances (e.g., the physical location of the victim row in a DRAM chip).}
{Although it might not be in the best interest of a DRAM manufacturer to make such understanding publicly available,\om{5}{\footnote{\om{5}{See~\cite{patel2022case} for a discussion and analysis of such issues.}}} \om{4}{rigorous} research in the public domain should continue to \om{5}{enable a much more detailed and rigorous} understanding of DRAM read disturbance. This is important because a better understanding of DRAM read disturbance among the broader research community enables the development of comprehensive solutions to the problem more quickly.}
Unfortunately, despite the existing research efforts expended towards understanding read disturbance{~\understandingRowHammerAllCitations{}}, scientific literature lacks 1)~rigorous experimental observations {on the \om{4}{\emph{spatial variation \om{5}{of} read disturbance}} in \om{4}{modern} DRAM chips and 2)~a concrete methodology \om{4}{for} leveraging this variation towards improving existing solutions \agy{4}{and crafting more effective attacks}.}}


{Our \emph{goal} in this paper is {to \om{4}{close this gap\om{5}{.} We aim to empirically analyze} the spatial variation \om{4}{of} read disturbance across DRAM rows and leverage this \om{4}{analysis} to improve existing solutions}. Doing so provides us {with} a deeper understanding of {the read disturbance in DRAM chips} to enable future research {on improving the effectiveness of existing {and future solutions}. We hope \agy{4}{and expect that our} analyses will pave the way for building \agy{4}{robust (i.e., reliable, secure, and safe) systems that mitigate DRAM read disturbance at low}
performance, energy, and area overheads 
\agy{4}{while DRAM chips become increasingly more vulnerable to read disturbance over generations.}

%% file: sections/04_methodology.tex
\section{Methodology}
\label{sec:methodology}

\copied{RowPress}{\om{4}{We describe} our DRAM testing infrastructure and the real DDR4 DRAM chips tested.}

\subsection{DRAM Testing Infrastructure}
\label{subsec:methodology_infra}

\begin{figure}[h]
\centering
\includegraphics[width=\linewidth]{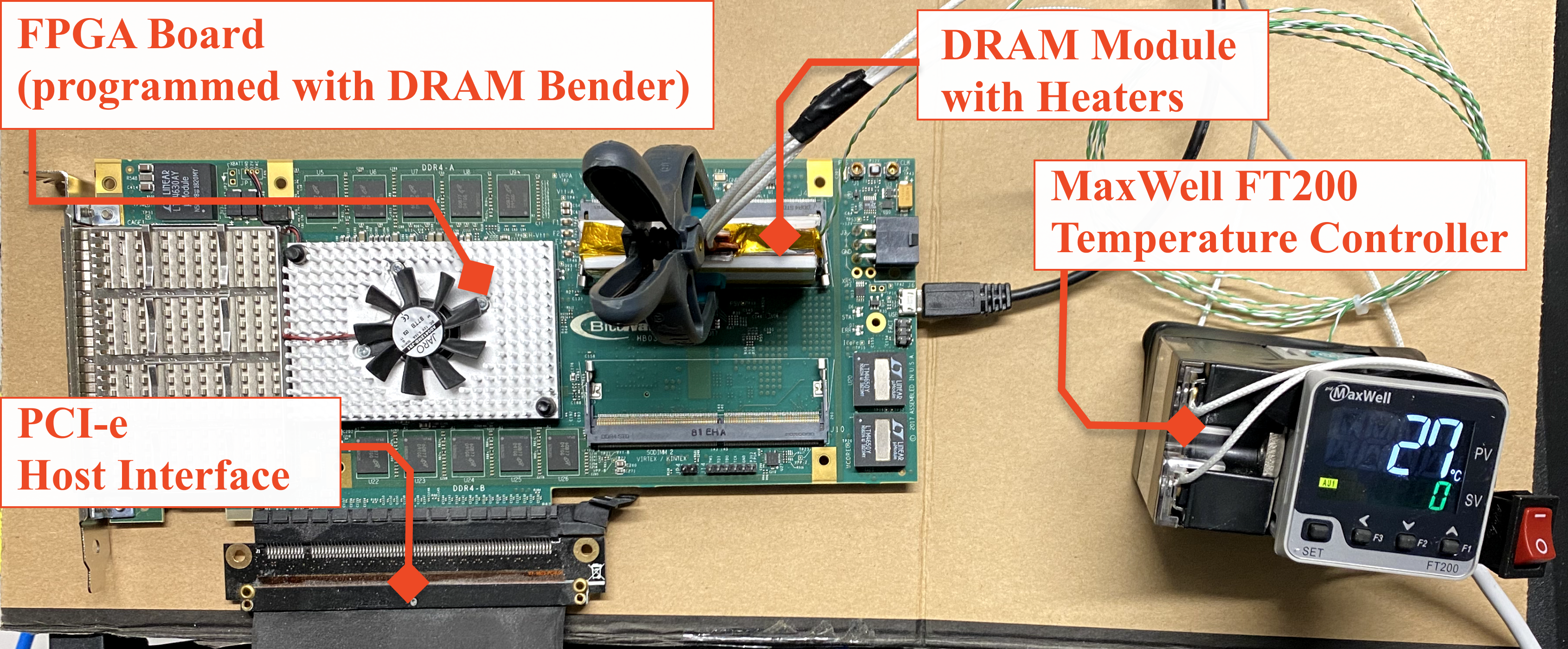}
\caption{Our DRAM Bender-based DDR4 DRAM testing infrastructure}
\label{fig:infrastructure}
\end{figure}

{{Fig.~\ref{fig:infrastructure} shows our FPGA-based DRAM testing infrastructure for testing} real DDR4 DRAM chips. {Our infrastructure} consists of four main components: 1)~an FPGA development board (Xilinx Alveo U200~\cite{alveo} for DIMMs or Bittware XUSP3S~\cite{xusp3s} for SODIMMs), programmed with DRAM Bender~\cite{olgun2022drambender, safari-drambender} to execute our test programs, 2)~a host machine that generates the test program and collects experimental results, 3)~a thermocouple temperature sensor and a pair of heater pads pressed against the DRAM chips that {heat} up the DRAM chips to a desired temperature, and 4)~a PID temperature controller (MaxWell FT200~\cite{maxwellFT200}) that controls the heaters and keeps the temperature at the desired level with a precision of $\pm$\SI{0.5}{\celsius}.}\footnote{{To evaluate temperature stability during RowHammer tests, we perform a double-sided RowHammer test with a \agy{5}{hammer} count of 1M and traverse across all rows in round-robin fashion for 24 hours at three different temperature levels. We sample the temperature \agy{4}{of three modules (one from each manufacturer)} every 5 seconds and observe a variation within the error margin of \SI{0.2}{\celsius}, \SI{0.3}{\celsius}, and \SI{0.5}{\celsius} at \SI{35}{\celsius}, \SI{50}{\celsius}, and \SI{80}{\celsius}, respectively.}}

\head{\om{4}{Eliminating} Interference Sources}
{To observe {read disturbance induced bitflips} in circuit level,
we \om{4}{eliminate}
{perform the best effort to eliminate} 
potential sources of interference 
\om{4}{to our best ability and control,}
{by taking {four} measures,} similar \om{4}{to the} methodology used by prior works~\cite{kim2020revisiting, orosa2021deeper, yaglikci2022understanding, hassan2021utrr, luo2023rowpress}.
First, we disable periodic refresh during the execution of our test programs to {prevent potential on-DRAM-die {TRR} mechanisms~\cite{frigo2020trrespass, hassan2021utrr} from refreshing victim rows so that we can} observe the DRAM chip's behavior at the circuit-level.
Second, we {strictly bound the execution time of} our test programs within {the} refresh window {of the tested DRAM chips at the tested temperature} to avoid data retention failures interfering with read disturbance failures.
{Third, we run each test ten times and record the
smallest (largest) observed \gls{hcfirst} (\gls{ber}) for each row across iterations to account for the worst-case.\footnote{{We observe a 5.7\% variation in the bit error rate across ten iterations.}}}
Fourth, we {verify} that the tested DRAM modules and chips have neither rank-level nor on-die ECC~\cite{patel2020beer, patel2021harp}.
{With these measures,} we directly observe and analyze all bitflips without interference.}

\vspace{-0.5em}
\subsection{Tested DDR4 DRAM Chips}
\vspace{-0.5em}
\label{subsec:methodology_dramchips}
\copied{RowPress}{Table~\ref{tab:dram_chip_list} shows the {\numchips{} real DDR4 DRAM chips (in \nummodules{} modules)} {spanning 10 different chip designs} that we test from all three major DRAM manufacturers. 
{To investigate whether our spatial variation analysis applies to different DRAM technologies, designs, and manufacturing processes,
we test various} DRAM chips with different \agy{7}{densities, die revisions, and chip organizations} from each DRAM chip manufacturer.}\footnote{{A DRAM chip's} technology node is {\emph{not} always} publicly available. 
We assume that two DRAM chips from the same manufacturer have the same technology node \emph{only} if they share both {1)~}the same die density and {2)~the same} die revision code.}

\input{tables/04_modules}

{To account for in-DRAM row address mapping~\cite{kim2014flipping, smith1981laser, horiguchi1997redundancy, keeth2001dram, itoh2013vlsi, liu2013experimental,seshadri2015gather, khan2016parbor, khan2017detecting, lee2017design, tatar2018defeating, barenghi2018software, cojocar2020rowhammer,  patel2020beer}, we reverse engineer the physical row address layout, following the prior works' methodology~\cite{kim2020revisiting, orosa2021deeper, yaglikci2022understanding, luo2023rowpress}.}

\subsection{DRAM Testing Methodology}

\head{Metrics}
{To characterize {a DRAM module's} vulnerability to read disturbance, we examine {the change in two metrics: \agy{5}{1)~\glsfirst{hcfirst}, where we count
each pair of activations to the two neighboring rows as
one hammer (e.g., one activation each to rows N – 1 and
N +1 counts as one hammer)~\cite{kim2020revisiting}, and 2)~\glsfirst{ber}}. A higher \gls{hcfirst} \om{5}{(\gls{ber}) indicates lower (higher)} vulnerability to read disturbance.}

\head{Tests}
Alg.~\ref{alg:test_alg} describes our core test loop \agy{4}{and two key functions we use: $hammer\_doublesided$ and $measure\_BER$}. 
{All our tests use \om{4}{the} double-sided hammering pattern \agy{4}{as specified in $hammer\_doublesided$ function and performed similarly by prior works}~\cite{kim2014flipping, kim2020revisiting, seaborn2015exploiting, orosa2021deeper, luo2023rowpress}. \agy{4}{$hammer\_doublesided$} {hammers two physically adjacent (i.e., aggressor) rows to a victim row \agy{4}{($RA_{victim}\pm1$)}} in an alternating manner. {In this context, one hammer is a pair of activations} to the two aggressor rows. \agy{4}{The $HC$ parameter in Alg.~\ref{alg:test_alg} defines the \agy{5}{hammer} count, i.e., the number of activations per aggressor row. The \gls{taggon} parameter in Alg.~\ref{alg:test_alg} defines the time an activated aggressor row remains open.}
We perform \agy{4}{the} double-sided {hammering} {in two different ways: 1)~}with the maximum activation rate possible within DDR4 command timing {specifications}~\cite{jedec2017ddr4,jedec2012ddr3} as this access pattern is stated as the most effective RowHammer {access pattern} on DRAM chips when {RowHammer solutions} are disabled~\cite{kim2014flipping, kim2020revisiting, frigo2020trrespass, cojocar2020rowhammer, seaborn2015exploiting, orosa2021deeper, olgun2023hbm}{; and 2)~with keeping aggressor rows open for longer \om{4}{than the minimum charge restoration time (\gls{taggon} > \gls{tras})} at each activation to \agy{4}{observe the effect of} RowPress, a recently demonstrated read disturbance phenomenon, \om{5}{which is different from} \agy{4}{RowHammer}~\cite{luo2023rowpress}.}}
\agy{4}{As $measure\_BER$ function (Alg.~\ref{alg:test_alg}) demonstrates, we initialize 1)~two aggressor rows and one victim row with opposite data patterns to \agy{5}{exacerbate read disturbance}~\cite{kim2020revisiting, kwong2020rambleed, cojocar2019eccploit, ji2019pinpoint}, 2)~perform double-sided hammer test, and 3)~read-back the data from the victim row and compare against the victim row's initial data pattern to calculate the bit error rate (BER).}
\agy{4}{Our core test loop sweeps different \gls{taggon} values, banks, and victim row addresses. 
First, We test three different \gls{taggon} values: 1)~\SI{36}{\nano\second} as the minimum \gls{tras} value, 2)~\SI{2}{\micro\second} as a large enough time window in which a streaming access pattern can fetch the whole content in the activated aggressor row, and 3)~\SI{0.5}{\micro\second} as a more realistic time window at which a DRAM row can remain open due to high row buffer hit rate~\cite{ghose2019demystifying, luo2023rowpress}.
Second, we sweep through banks 1, 4, 10, and 15 as representative banks from each bank group~\cite{jedec2017ddr4, olgun2023drambender, safari-drambender}. 
Third, we test \emph{all} rows in a tested bank using 14 different \agy{5}{hammer} counts and six different data patterns.}

\head{Hammer Counts} We conduct our tests by using a set of \agy{5}{hammer} counts on all DRAM rows instead of finding \gls{hcfirst} precisely for each row. \agy{4}{This is because}
\gls{hcfirst} significantly varies across rows, and thus, causes a large experiment time (e.g., several weeks or even months) to find \gls{hcfirst} at high precision (e.g., \om{4}{within} $\pm 10$ hammers) for each row individually. 
Therefore, we test the DRAM chips under 14 different hammer counts from $1K$ to $128K$ as specified in Alg.~\ref{alg:test_alg}.\footnote{{$K$ is $2^{10}$ (\emph{not} $10^3$) unless otherwise specified.}}  

\head{{Data Patterns}} {{We use} six commonly used data patterns~\cite{chang2016understanding,chang2017understanding,khan2014efficacy,khan2016parbor,khan2016case,kim2020revisiting,lee2017design,mukhanov2020dstress,orosa2021deeper, kim2014flipping, liu2013experimental}: {row stripe, checkerboard, column stripe, and the opposites of these three data patterns that are shown in \tabref{tab:data_patterns} in detail}. We identify the worst-case data pattern ($WCDP$) for each row {as the data pattern that results in the largest \gls{ber} at the \agy{5}{hammer} count of $128K$}.\footnote{We find that a \agy{5}{hammer} count of 128K is both 1)~low enough to be used in a system-level attack in a real system~\cite{frigo2020trrespass}, and 2)~high enough to provide a large number of bitflips in \emph{all} DRAM modules we tested{.}}}
\agy{4}{Then, we sweep the \agy{5}{hammer} count from 1K to 96K and measure \gls{ber} for the WCDP of each row.}

\begin{table}[h]
    \centering
    \footnotesize
    \caption{Data patterns used in our tests}
    \begin{tabular}{l|cc}
     \bf{Data Pattern}   & \bf{{Aggressor Rows}} & \bf{{Victim Row}}\\
    \hline
    \hline
    {Row Stripe (RS)}         & $0xFF$ & $0x00$ \\
    {Row Stripe Inverse (RSI)} & $0x00$ & $0xFF$ \\
    \hline
    {Column Stripe (CS)}         & $0xAA$ & $0xAA$ \\
    {Column Stripe Inverse (CSI)} & $0x55$ & $0x55$ \\
    \hline
    {Checkerboard (CB)}         & $0xAA$ & $0x55$ \\
    {Checkerboard Inverse (CBI)} & $0x55$ & $0xAA$ \\
    \hline
    \hline
    
    \end{tabular}
    \label{tab:data_patterns}
\end{table}


\SetAlFnt{\scriptsize}
\RestyleAlgo{ruled}
\begin{algorithm}
\caption{Test for profiling the spatial variation of read disturbance in DRAM}\label{alg:test_alg}
  \DontPrintSemicolon
  \SetKwFunction{FVPP}{set\_vpp}
  \SetKwFunction{FMain}{test\_loop}
  \SetKwFunction{FHammer}{measure\_$BER$}
  \SetKwFunction{initialize}{initialize\_row}
  \SetKwFunction{initializeaggr}{initialize\_aggressor\_rows}
  \SetKwFunction{measureber}{measure\_BER}
  \SetKwFunction{FMeasureHCfirst}{measure\_\gls{hcfirst}}
  \SetKwFunction{compare}{compare\_data}
  \SetKwFunction{Hammer}{hammer\_doublesided}
  \SetKwFunction{Gaggressors}{get\_aggressors}
  \SetKwFunction{FWCDP}{get\_WCDP}
  \SetKwProg{Fn}{Function}{:}{}

  \tcp{$RA_{victim}$: Victim row address}
  \tcp{$WCDP$: Worst-case data pattern \om{4}{for the victim row}}
  \tcp{$HC$: \agy{5}{Hammer Count: number of activations} per aggressor row}
  \tcp{$ACT$: Row activation command to open a DRAM row}
  \tcp{$PRE$: Precharge command to close a DRAM row}
  \tcp{$WAIT$: Wait for the specified amount of time}
  \tcp{$t_{AggOn}$: Aggressor row on time}
  \tcp{$t_{RP}$: Precharge latency timing constraint}
  \Fn{\Hammer{{$RA_{victim}$}, $HC$, $t_{AggOn}$}}{
        \While{$i < HC$}{
            ACT({$RA_{victim}+1$})\;
            WAIT({$t_{AggOn}$})\;
            PRE({})\;
            WAIT({$t_{RP}$})\;
            ACT({$RA_{victim}-1$})\;
            WAIT({$t_{AggOn}$})\;
            PRE({})\;
            WAIT({$t_{RP}$})\;
            i++\;
        }
  }\;

  \Fn{\FHammer{$RA_{victim}$, $WCDP$, $HC$, $t_{AggOn}$}}{
        \initialize($RA_{victim}$, $WCDP$)\;
        {\initializeaggr($RA_{victim}$, {bitwise\_inverse($WCDP$)})}\;
        \Hammer({$RA_{victim}$}, $HC$, $t_{AggOn}$)\;
        $BER{_{row}} =$ \compare($RA_{victim}$, $WCDP$)\;
        \KwRet $BER{_{row}}$\;
  }\;

  \Fn{\FMain{}}{
    \ForEach{$t_{AggOn}$ in [36ns, 0.5us, 2us]}{
        \ForEach{$Bank$ in [1, 4, 10, 15]}{
            \ForEach{$RA_{victim}$ in $Bank$}{
                \tcp{Find the worst-case data pattern}
                \ForEach{$DP$ in [RS, RSI, CS, CSI, CB, CBI]}{
                    \FHammer{$RA_{victim}$, $DP$, 128K, $t_{AggOn}$}\;
                    WCDP = DP that causes largest BER\;
                }
    
                \tcp{Sweep the hammer count using WCDP}
                \ForEach{$HC$ in [1,2,4,8,12,16,24,32,40,48,56,64,96]K}{
                    \FHammer{$RA_{victim}$, $WCDP$, $HC$, $t_{AggOn}$}\;
                }    
            }
        }
    }
  }
\end{algorithm}

\head{Finding Physically Adjacent Rows} \copied{HammerDimmer}{DRAM-internal address mapping schemes~\cite{cojocar2020rowhammer, salp} are used by DRAM manufacturers to translate {\emph{logical}} DRAM addresses (e.g., row, bank, {and} column) that are exposed over the DRAM interface (to the memory controller) to physical {DRAM} addresses {(e.g., physical location of a row)}. {Internal address mapping schemes allow 
{1)~}post-manufacturing row repair techniques to repair erroneous DRAM rows by remapping \om{4}{such} rows to spare rows and 
{2)~}DRAM manufacturers organize DRAM internals in a cost-optimized way, e.g., by organizing internal DRAM buffers hierarchically~\cite{khan2016parbor,vandegoor2002address}.} The mapping scheme can substantially vary across different DRAM {chips}~\cite{barenghi2018software,cojocar2020rowhammer,horiguchi1997redundancy,itoh2013vlsi,keeth2001dram,khan2016parbor,khan2017detecting,kim2014flipping,lee2017design,liu2013experimental,patel2020beer,orosa2021deeper,saroiu2022price,patel2022case}. For every victim DRAM row {that we test}, we identify the two {neighboring physically-adjacent} DRAM row addresses that the memory controller can use to access the {aggressor} rows in a double-sided RowHammer attack. To do so, we reverse-engineer the physical row organization {using} techniques described in prior work{s}~\cite{kim2020revisiting, orosa2021deeper}.}

\pagebreak
\head{Temperature} We maintain the DRAM chip temperature at \SI{80}{\celsius}, \om{4}{which is} very close to {the maximum point of the} normal operating condition of \SI{85}{\celsius}~\cite{jedec2017ddr4}. {We choose this temperature because prior works show that increasing temperature tends to reduce DRAM chips' overall realiability~\cite{liu2013experimental, orosa2021deeper, orosa2022spyhammer, luo2023rowpress}.}{\footnote{{Prior works~\cite{orosa2021deeper, luo2023rowpress} \om{4}{demonstrate} a complex interaction between temperature and a row’s read disturbance \om{4}{(especially RowHammer)} vulnerability and suggest that each DRAM chip should be tested at all temperature levels to account for the effect of temperature. Thus, fully understanding the effects of temperature and aging requires extensive characterization studies, requiring many months-long\ominlinecomment{4}{correct? Giray:unfortunately, yes} testing time. Therefore, we leave such studies for future work.}}} {Due to time and space limitations, we leave a rigorous characterization of temperature's effect for future work, while presenting the preliminary analysis \agy{4}{where}} {we repeat double-sided RowHammer tests at \SI{50}{\celsius} on 5K randomly selected DRAM rows at nine different \agy{5}{hammer} counts. We observe that the variation in overall \gls{ber} with the effect of temperature is less than 0.5\%.}

%% file: tables/04_modules.tex
\begin{table}[h!]
  \centering
  \footnotesize
  \caption{Tested DDR4 DRAM Chips.}
    \begin{tabular}{l|lrlcc}
        &{\bf DIMM}&{\bf \# of}&{\bf Density}&{\bf Chip}&{{\bf Date}}\\
        {{\bf Mfr.}} & \textbf{ID} & {{\bf Chips}}  & {{\bf Die Rev.}}& {{\bf Org.}}& {{\bf (ww-yy)}}\\
        \hline 
        \hline 
        \multirow{3}{*}{\begin{tabular}[c]{@{}l@{}}Mfr. H \\ (SK Hynix)\end{tabular}} & H0 &  $8$ & 16Gb -- A   & x8  & 51-20 \\     
        & H1, H2, H3 &  $3\times8$ & 16Gb -- C   & x8  & 48-20   \\   
        & H4 &  $8$ &  8Gb -- D   & x8  & 48-20   \\
        \hline
                 & M0 & $4$  & 16Gb -- E   & x16 & 46-20  \\
        Mfr. M   & M1, M3 & $2\times16$ & 8Gb  -- B   & x4  & N/A   \\
        (Micron) & M2 & $16$ & 16Gb -- E   & x4  & 14-20  \\  
                 & M4 & $4$  & 16Gb -- B   & x16 & 26-21  \\
        \hline 
                  & S0, S1 & $2\times8$ & 8Gb  -- B   & x8  & 52-20  \\
        Mfr. S    & S2 &  $8$ & 8Gb  -- B   & x8  & 10-21  \\
        (Samsung) & S3 &  $8$ & 4Gb  -- F   & x8  & N/A   \\
                  & S4 & $16$ & 8Gb  -- C   & x4  & 35-21  \\
        \hline
        \hline
    \end{tabular}
    \label{tab:dram_chip_list}
\end{table}

%% file: sections/05_characterization.tex
\section{Spatial Variation in DRAM Read Disturbance}
\label{sec:characterization}
{This section presents the first {rigorous} {spatial variation} analysis of read disturbance across DRAM rows. Many prior works~\cite{kim2014flipping, kim2020revisiting, park2014active, park2016experiments, park2016statistical} analyze RowHammer vulnerability {at the} DRAM bank granularity across many DRAM modules without {providing analysis of} the variation of this vulnerability across rows.} {Recent works~\cite{orosa2021deeper, yaglikci2022understanding, luo2023rowpress, olgun2023hbm} analyze the variation in RowHammer vulnerability across DRAM rows. However, these analyses are limited to a small subset of DRAM rows (4K to 9K), while a DRAM bank typically has $>16K$ DRAM rows~\cite{jedec2017ddr4, datasheetM393A2K40CB2-CTD, datasheetK4A8G085WB-BCTD, datasheetH5ANAG8NCJR-XN, datasheetH5AN8G8NDJR-XNC, datasheetMTA4ATF1G64HZ-3G2E1, datasheetMT40A1G16KD-062E, datasheetMTA18ASF2G72PZ-2G3B1QK, datasheetMTA36ASF8G72PZ-2G9E1TI, datasheetMTA4ATF1G64HZ-3G2B2}. Thus, \om{4}{prior} works might \emph{not} fully reflect the vulnerability profile of real DRAM chips. Fully \om{4}{characterizing and understanding} the vulnerability profile is crucial to \om{4}{avoiding} read disturbance bitflips as existing \om{4}{read disturbance} solutions must be properly configured \om{4}{based on proper characterization}~\cite{kim2014flipping, kim2014architectural, yaglikci2021blockhammer, park2020graphene, qureshi2022hydra, saxena2022aqua, saileshwar2022randomized, bostanci2024comet, olgun2024abacus}.} This section presents a more rigorous and targeted read disturbance characterization study of \numchips{} real DDR4 DRAM chips {spanning \om{5}{\param{10}} different \agy{7}{chip designs}}, following the methodology described in \secref{sec:methodology}.

\subsection{Bit Error Rate Across DRAM Rows}
\label{sec:char_ber}
\begin{figure}[h!]
    \centering
    \includegraphics[width=\linewidth]{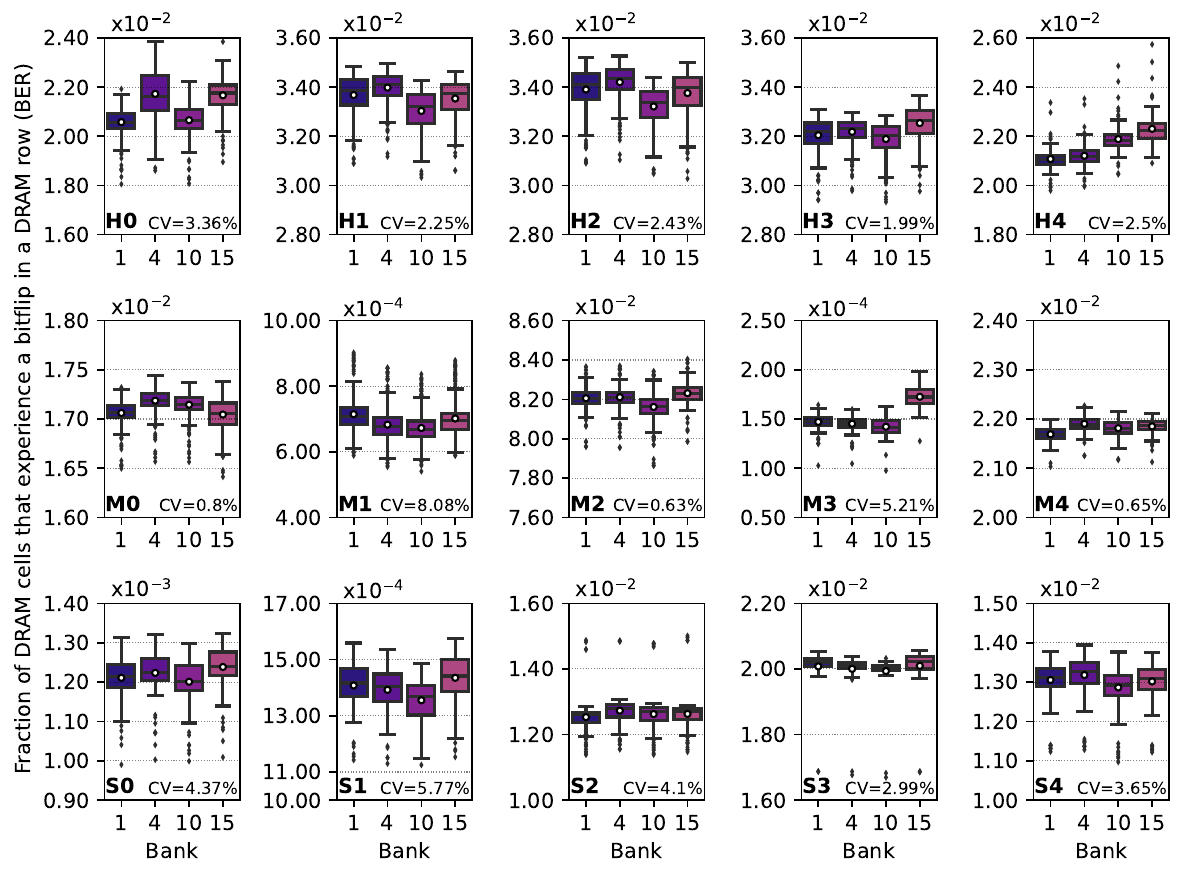}
    \caption{Distribution of \gls{ber} across DRAM rows and bank groups}
    \label{fig:ber_hist}
\end{figure}

We investigate the variation in the number of bitflips caused by read disturbance for \om{4}{a} \agy{5}{hammer} count of 128K and \gls{taggon} of \SI{36}{\nano\second}. \figref{fig:ber_hist} shows the distribution of observed \gls{ber} for each DRAM row across all tested DRAM banks and modules from three main manufacturers in a box-and-whiskers plot.\footnote{\label{fn:boxplot}{{The box is lower-bounded by the first quartile (i.e., the median of the first half of the ordered set of data points) and upper-bounded by the third quartile (i.e., the median of the second half of the ordered set of data points).
The \gls{iqr} is the distance between the first and third quartiles (i.e., box size).
Whiskers mark the central 1.5\gls{iqr} range, and white circles show the mean values.}}} Each of the three rows of subplots is dedicated to modules from a different manufacturer, and each subplot shows data from a different DRAM module. The x-axis shows the bank address and the y-axis shows the \glsfirst{ber}. \agy{4}{We annotate each module's name and variation across rows and banks in terms of the coefficient of variation (CV)\footnote{Coefficient of variation is the standard deviation of a distribution, normalized to the mean~\cite{faber2012statistics, brian1998dictionary}.} at the bottom of each subplot.} We make Obsvs.~\ref{obsv:ber_across_rows}-\ref{obsv:ber_across_modules} from \figref{fig:ber_hist}.

\pagebreak
\observation{\gls{ber} varies across DRAM rows in a DRAM module.\label{obsv:ber_across_rows}}
For example, DRAM rows in \om{4}{modules} M1 and S1 exhibit coefficient of variations (CV) of 8.08\% and 5.77\%, respectively, on average across all tested banks. 

\observation{Different banks within the same DRAM module exhibit similar \gls{ber} to each other.\label{obsv:ber_across_banks}} 
As the box plots for different banks largely overlap with each other in the y-axis, we observe a smaller variation in \gls{ber} across banks compared to across rows in a bank for all tested modules except H4 and M3. 
For example, the average (minimum/maximum) \gls{ber} across all DRAM rows in four different banks of M0 are 1.71\% (1.65\%/1.73\%), 1.71\% (1.66\%/1.74\%), 1.70\% (1.64\%/1.74\%), and 1.72\% (1.66\%/1.74\%).



\observation{\gls{ber} can significantly vary across different DRAM modules from the same manufacturer.\label{obsv:ber_across_modules}}
For example, modules M0, M1, and \agy{4}{M3} show \gls{ber} distributions that \agy{4}{are strictly distinct from each other} across their mean \gls{ber} values.
From Obsvs.~\ref{obsv:ber_across_rows}-\ref{obsv:ber_across_modules}, we draw Takeaway~\ref{take:ber_across_everything}. 
\takebox{\gls{ber} significantly varies across different DRAM rows within a DRAM bank and across different DRAM modules, while different banks in a DRAM module exhibit similar \gls{ber} distributions \om{4}{to} each other.\label{take:ber_across_everything}} 

\begin{figure}[h!]
    \centering
    \includegraphics[width=\linewidth]{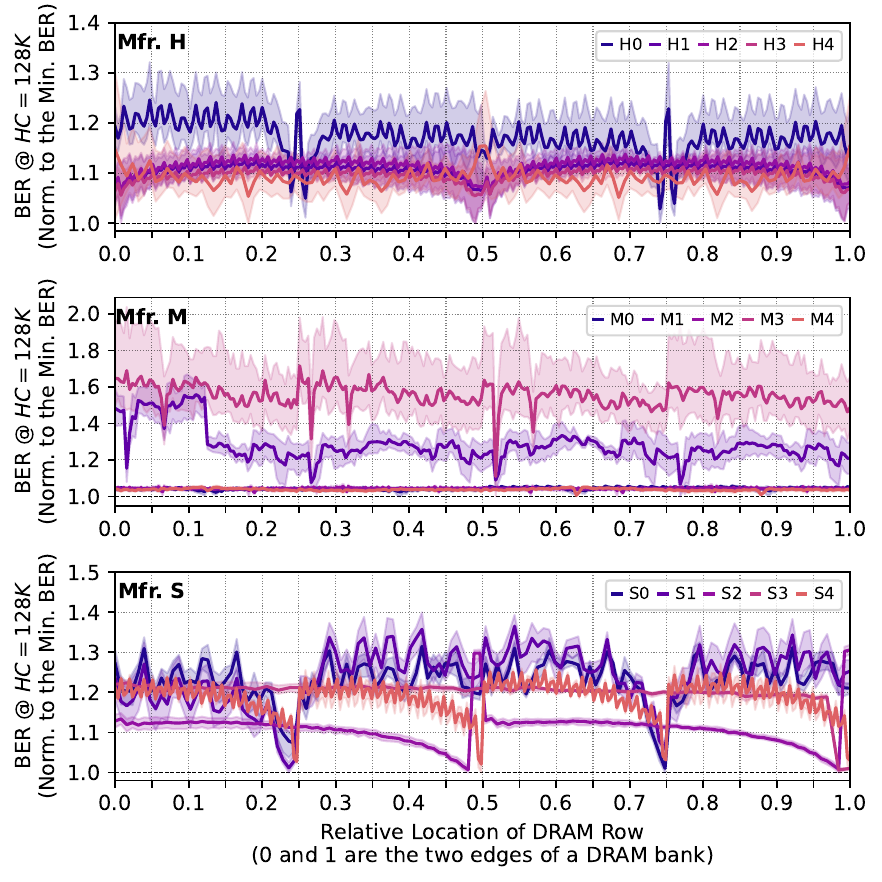}
    \caption{Distribution of \gls{ber} across DRAM rows}
    \label{fig:ber_vs_row}
\end{figure}

To understand the spatial variation of rows with high and low \gls{ber}s, we analyze their locations within their banks. \figref{fig:ber_vs_row} shows \agy{4}{how \gls{ber} varies} as the row address increases. \agy{4}{The x-axis shows a DRAM row's relative location in its bank, where 0.0 and 1.0 are the two edges of a DRAM bank. The y-axis shows the \gls{ber} the corresponding DRAM row experiences at a \agy{5}{hammer} count of 128K, normalized to the minimum \gls{ber} observed across all rows in all tested banks in a module.} Each subplot is dedicated to a different manufacturer, and each curve represents a different DRAM module. The shades around the curves show the minimum and maximum values for a given row address across different DRAM banks in a module. 
We make Obsvs.~\ref{obsv:ber_row_repeating_patterns} and~\ref{obsv:ber_across_chunks} from \figref{fig:ber_vs_row}.


\observation{\gls{ber} repeatedly increases and decreases with \agy{4}{different intervals of row distances in different DRAM modules}.\label{obsv:ber_row_repeating_patterns}}
 For example, \gls{ber} curve of S4 follows a repeatedly increasing and decreasing pattern across all rows, where it shows local minimums at 0.25, 0.50, 0.75, and 1.00.
\agy{4}{We hypothesize that this regularity in \gls{ber} variation can be caused by design decisions (design-induced variation), e.g., row's distance from subarray boundaries and I/O circuitry, as discussed by prior works~\cite{lee2017design, orosa2021deeper, olgun2023hbm}.}

\observation{Average \gls{ber} can vary across large chunks of a DRAM bank. \label{obsv:ber_across_chunks}}
For example, the average (minimum/maximum) normalized \gls{ber} \agy{4}{in the module M1} across DRAM rows between \om{4}{relative locations} 0.03 and 0.12 is 1.51 (1.31 /1.67 ) while it is 1.25 (1.00 /1.42 ) between \om{4}{relative locations} 0.20 and 1.00.
This \agy{4}{discrepancy in \gls{ber} across large chunks of rows} does \emph{not} consistently occur across all tested modules. 
Understanding the root cause of this discrepancy 
requires extensive knowledge and insights into the circuit design and manufacturing process of the particular DRAM modules exhibiting this behavior. Unfortunately, this piece of information is proprietary and \emph{not} publicly disclosed by the manufacturers. \agy{4}{We hypothesize that the root cause of this discrepancy can be the variation in the manufacturing process, leading to a part of DRAM chip being more vulnerable to read disturbance compared to other parts.}
From Obsvs.~\ref{obsv:ber_row_repeating_patterns} and~\ref{obsv:ber_across_chunks}, we derive Takeaway~\ref{take:ber_regular}.

\takebox{\agy{4}{\gls{ber} values in a DRAM bank exhibit repeating patterns as DRAM row address increases}, and certain chunks of rows can exhibit higher \gls{ber} than the rest of the rows.\label{take:ber_regular}}

\subsection{Minimum Activation Count to Induce a Bitflip}
\label{sec:char_hcfirst}

\begin{figure}[h!]
    \centering
    \includegraphics[width=\linewidth]{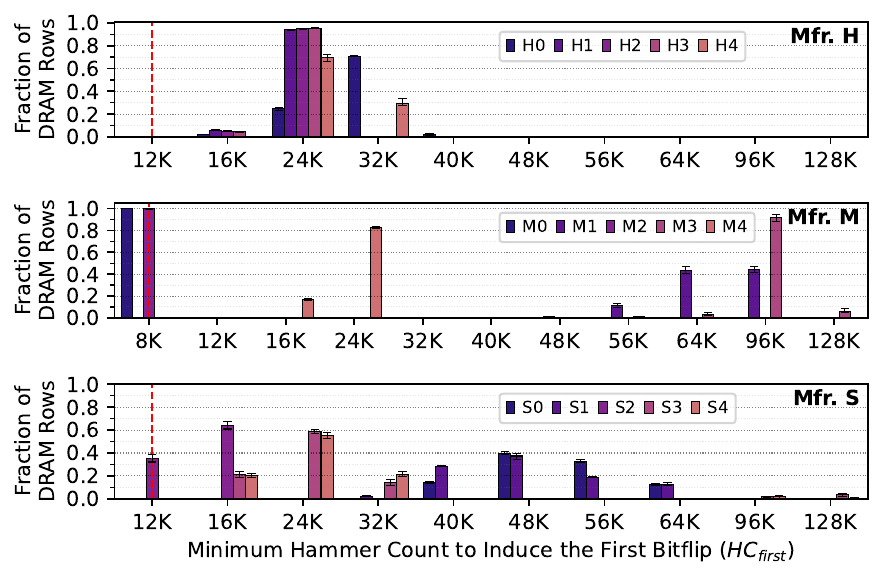}
    \caption{Distribution of \gls{hcfirst} across DRAM rows}
    \label{fig:hcfirst_hist}
\end{figure}

We investigate the variation in \gls{hcfirst} across DRAM rows.
To do so, we repeat our tests at \param{14} different \agy{5}{hammer} counts from 1K to 128K (Algorithm~\ref{alg:test_alg}). We define \agy{6}{a row's \gls{hcfirst}} as the minimum of the tested \agy{5}{hammer} counts at which the row experiences a bitflip. \figref{fig:hcfirst_hist} shows the distribution of \gls{hcfirst} values across rows. Each subplot shows the distribution for a different manufacturer. The x- and y-axes show the \gls{hcfirst} values and the fraction of the DRAM rows with the specified \gls{hcfirst} value, respectively. Different colors represent different modules. The error bars mark the minimum/maximum of a given value across tested banks. Red vertical dashed line marks the minimum \gls{hcfirst} that we observe across all rows in tested modules from a manufacturer. We make Obsvs.~\ref{obsv:hcfirst_rows}--\ref{obsv:hcfirst_modules} \agy{4}{from \figref{fig:hcfirst_hist}}.

\observation{\gls{hcfirst} values significantly vary across DRAM rows but not across banks.\label{obsv:hcfirst_rows}} For example, S0 and S1 contain rows that experience bitflips \agy{4}{at \agy{5}{hammer} counts of 32K and 24K, respectively, while they also have rows} that do not experience bitflips until 128K. Despite this large variation, the variation across banks is significantly low, as error bars show.


\observation{Different DRAM modules from the same manufacturer can exhibit significantly different \gls{hcfirst} distributions.\label{obsv:hcfirst_modules}} For example, rows from M0 and M4 exhibit \gls{hcfirst} values from 8K to 40K and 12K to 96K, respectively. 

\takebox{\gls{hcfirst} varies \om{4}{significantly} across different DRAM rows within a DRAM bank and across different DRAM modules, while different banks in a DRAM module exhibit similar \gls{hcfirst} distributions with each other.\label{take:hcfirst_across_everything}} 

\begin{figure}[h!]
        \centering
        \includegraphics[width=\linewidth]{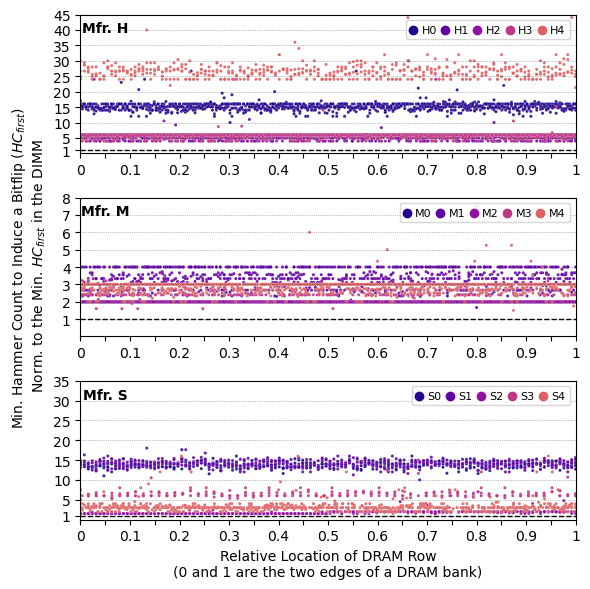}
        \caption{Distribution of \gls{hcfirst} across DRAM rows}
        \label{fig:hcfirst_vs_row}
    \end{figure}

To understand the spatial variation in \gls{hcfirst} across rows, we investigate \agy{4}{how a row's \gls{hcfirst} changes with the row's location within the DRAM bank. \figref{fig:hcfirst_vs_row} shows the row's relative location on the x-axis and its \gls{hcfirst}, normalized to the minimum \gls{hcfirst} observed in the corresponding module.}
Each subplot corresponds to a different manufacturer, and different modules are color-coded. We make Obsvs.~\ref{obsv:hcfirst_everything_everywhere} and \ref{obsv:hcfirst_row_repeating_patterns} from \figref{fig:hcfirst_vs_row}.

\observation{\agy{4}{\gls{hcfirst} values vary significantly across rows.}\label{obsv:hcfirst_everything_everywhere}}
For example, the \agy{4}{module H0 exhibits \gls{hcfirst} values that are between $8\times$ and $20\times$ the minimum \gls{hcfirst} observed in the bank between relative row addresses 0.02 and 0.03.}

\observation{Variation in \gls{hcfirst} does not exhibit a \agy{4}{regular} trend as the row address increases.\label{obsv:hcfirst_row_repeating_patterns}} \agy{4}{For example, the data points of modules H4 \agy{5}{concentrate at the y-axis values of $24\times$ and $32\times$ across \emph{all} rows in a bank with \emph{no} regular transition pattern across them}.} \om{4}{This observation \om{5}{is discrepant with} Obsv.~\ref{obsv:ber_row_repeating_patterns} we have for \gls{ber}.} \agy{4}{\om{5}{The} discrepancy across Obsvs.~\ref{obsv:ber_row_repeating_patterns} and~\ref{obsv:hcfirst_row_repeating_patterns} shows that although read disturbance vulnerability varies regularly across rows in terms of the fraction of DRAM cells experiencing bitflips, \agy{5}{the \gls{hcfirst} values across the weakest DRAM cells do \emph{not} exhibit such a regular variation pattern}.} 
From Obsvs.~\ref{obsv:hcfirst_everything_everywhere} and \ref{obsv:hcfirst_row_repeating_patterns}, we derive Takeaway~\ref{take:hcfirst_across_rows}.

\takebox{\gls{hcfirst} varies \agy{4}{significantly and irregularly across rows and banks in a DRAM module.}\label{take:hcfirst_across_rows}}

\subsection{Effect of RowPress}
\label{sec:char_access_pattern}
\begin{figure}[h!]
    \centering
    \includegraphics[width=\linewidth]{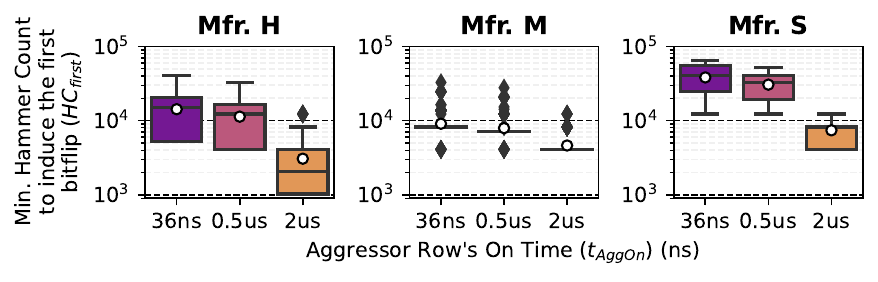}
    \caption{\om{4}{Effect of \gls{taggon} on \gls{hcfirst}}}
    \label{fig:hcfirst_tAggOn}
\end{figure}

We analyze the effect of the recently discovered read disturbance phenomenon, RowPress~\cite{luo2023rowpress}, on the \gls{hcfirst} distribution. To do so, we repeat our tests with \gls{taggon} configurations of \SI{0.5}{\micro\second} and \SI{2}{\micro\second} instead of \SI{36}{\nano\second}.\footnote{We choose these \gls{taggon} values because they are large enough to show the effects of RowPress and realistic, such that an adversarial access pattern can easily force these \gls{taggon} values by accessing different cachelines in a DRAM row. We do \emph{not} sweep all possible \gls{taggon} values due to the limitations in experiment time.}
\agy{4}{\figref{fig:hcfirst_tAggOn} shows 
a box-and-whiskers plot\footref{fn:boxplot} of the \gls{hcfirst} distribution across all rows in all tested modules under the three different \gls{taggon} values we test. The x-axis shows the \gls{taggon} values, and the y-axis shows the \gls{hcfirst} values. Different subplots show modules from different manufacturers.
We make Obsvs.~\ref{obsv:hcfirst_taggon_decreasing} and~\ref{obsv:hcfirst_taggon_heterogeneity} from \figref{fig:hcfirst_tAggOn}.}
\agycomment{4}{We will add the other scatter plot to the extended version.}

\observation{\gls{hcfirst} decreases with increasing \gls{taggon} for the vast majority of DRAM rows.\label{obsv:hcfirst_taggon_decreasing}} 
\agy{4}{We observe that both mean values and the box (\gls{iqr}) boundaries decrease on the y-axis when \gls{taggon} increases on the x-axis.}

\observation{\gls{hcfirst} values vary significantly across DRAM rows even when \gls{taggon} is \SI{2}{\micro\second}.\label{obsv:hcfirst_taggon_heterogeneity}}
\agy{4}{For example, \gls{hcfirst} distribution across rows in module H2 exhibits the coefficient of variation (CV) values of 25.0\%, 23.0\%, and 30.4\% for \gls{taggon} values of \SI{36}{\nano\second}, \SI{0.5}{\micro\second}, and \SI{2}{\micro\second}.}

From Obsvs.~\ref{obsv:hcfirst_taggon_decreasing} and~\ref{obsv:hcfirst_taggon_heterogeneity}, we draw Takeaway~\ref{take:taggon_hcfirst}.

\takebox{\agy{4}{\gls{hcfirst} values reduce as \gls{taggon} increases and vary significantly across rows for large \gls{taggon} values (e.g., \SI{2}{\micro\second}).}\label{take:taggon_hcfirst}}

\subsection{Spatial Features}
\label{sec:char_feature_selection}
This section investigates potential correlations between a DRAM row's vulnerability to read disturbance and the row's spatial features. To do so, we consider a set of features that might affect a DRAM row's reliable operation based on the findings of prior works~\cite{aldram, chang_understanding2017, chang2016understanding, lee2017design, kim2018solar, orosa2021deeper}\omcomment{4}{be more comprehensive}: 1)~bank address, 2)~row address, 3)~subarray address, 4)~row's distance to the sense amplifiers, i.e., subarray boundaries.
To perform this analysis, subarray boundary identification is critical. Unfortunately, this information is \emph{not} publicly available. To address this problem, we reverse-engineer the subarray boundaries. 

\subsubsection{Subarray Reverse Engineering}
\label{sec:char_subarray_reverse_engineering}
We leverage two key insights.

\begin{figure}[h!]
    \centering
    \includegraphics[width=\linewidth]{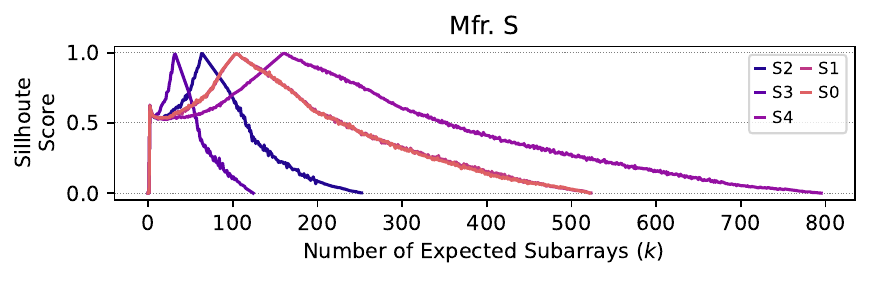}
    \caption{Silhouette score of classification of DRAM rows into subarrays using \om{4}{the} k-means algorithm}
    \label{fig:silhouette_score}
\end{figure}

\head{Key Insight 1}
First, a \om{4}{row located at a subarray boundary} can be disturbed by hammering or pressing its neighboring rows \emph{only} on one side of the row instead of both sides. \om{4}{Exploiting} this observation, we cluster the DRAM rows based on row address and the number of rows that single-sided hammering or pressing a given row affects.
We do \agy{4}{so} using \om{4}{the} k-means clustering algorithm\cite{kmeans}. Because the number of subarrays is initially unknown, we sweep the parameter k and choose the best k value based on the clustering's silhouette score~\cite{rousseeuw1987silhouettes}. 
As a representative example, \figref{fig:silhouette_score} shows the silhouette score of the classification of DRAM rows into subarrays using the k-means algorithm. We sweep the parameter k on the x-axis and show the silhouette score on the y-axis. \agy{4}{Different curves represent different modules from Mfr. S.}

From \figref{fig:silhouette_score}, we observe that the silhouette score reaches a global maximum and decreases monotonically as we sweep the k-parameter. Based on this observation, we hypothesize that the k-value at the global maximum is the number of subarray boundaries in a DRAM bank, and each cluster for this k-value is a subarray containing the rows in the cluster.

\head{Key Insight 2} Since DRAM rows share a local bitline within a subarray, it is possible to copy one row's (i.e., source row) data to another row (i.e., destination row) within the same subarray (i.e., also known as \om{4}{the intra-subarray} RowClone operation~\cite{seshadri2013rowclone}). Prior works~\cite{yaglikci2022hira, gao2019computedram, gao2022frac, olgun2021pidram, yuksel2023pulsar, yuksel2024functionallycomplete} already show that it is possible to perform RowClone in off-the-shelf DRAM chips by violating timing constraints \om{4}{such that two rows are activated in quick succession}.
We conduct RowClone tests following the prior work's methodology~\cite{gao2019computedram}. \agy{4}{If the source row's content is successfully copied to the destination row with \emph{no} bitflips, both the source and destination rows have to be in the same subarray. However, the opposite case (an unsuccessful RowClone operation) does \emph{not} necessarily mean that the two rows are in different subarrays. This is because intra-subarray RowClone is \emph{not} officially supported \om{5}{in existing DRAM chips}, and thus \emph{not} guaranteed to work reliably across all rows in a subarray.}

\agy{4}{We first identify the candidates of subarray boundaries using \om{5}{Key Insight 1} based on the single-sided RowHammer tests. Second, we test these subarray boundaries using \om{5}{Key Insight 2} based on the intra-subarray RowClone tests such that a successful intra-subarray RowClone operation invalidates a candidate subarray boundary since the source and the destination rows have to be in the same subarray, and thus there \emph{cannot} be a subarray boundary in between those two rows.}
Our analysis identifies differently sized subarrays (from 330 to 1027 rows per subarray) and different numbers of subarrays (from 32 to 206 subarrays per bank) across the tested chips. Unfortunately, we do \emph{not} have the ground truth design to verify our results.

\subsubsection{Correlation Analysis}
\label{sec:spatial_features_correlation_analysis}
We \agy{4}{analyze} the correlation \agy{4}{between a DRAM row's} spatial features and \agy{4}{the row's \gls{hcfirst} value.}
\agy{4}{As spatial features,} we take each bit in the binary representation of \agy{4}{a DRAM row's} four properties: 1)~bank address, 2)~row address, 3)~subarray address, and 4)~row's distance to the sense amplifiers. 
\agy{4}{We use each of the spatial features for each DRAM row to predict the row's \gls{hcfirst} among 14 tested \agy{5}{hammer} counts}. 
We compare the prediction and real experiment results to create the confusion matrix~\cite{aalst2010processmining} and calculate the F1 score~\cite{aalst2010processmining} for each feature.

\begin{figure}[h!]
    \centering
    \includegraphics[width=\linewidth]{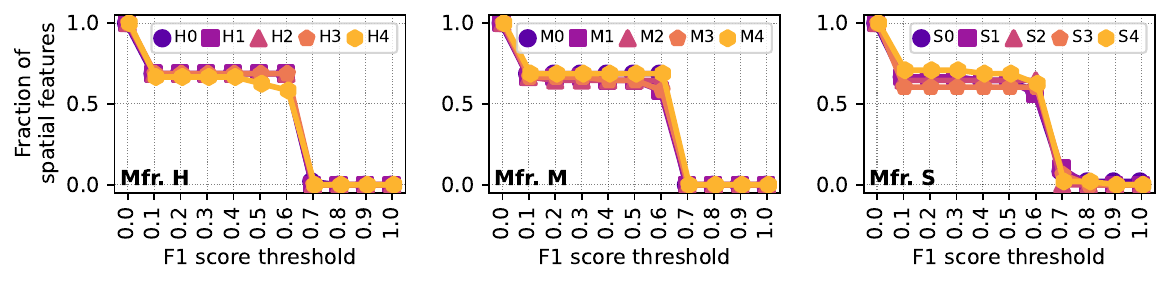}
    \caption{Fraction of spatial features vs F1 score threshold}
    \label{fig:hcfirst_correlations}
\end{figure}

\agy{4}{\figref{fig:hcfirst_correlations} shows the fraction of spatial features that \emph{strongly correlate} with the row's \gls{hcfirst}. \agy{5}{We consider a spatial feature's correlation with  \gls{hcfirst} \om{6}{to be} stronger if predicting \gls{hcfirst} based on the spatial feature results in a larger F1 score.}
The x-axis sweeps the F1 score threshold from 0 to 1, and the y-axis shows the fraction of spatial features that correlate with \gls{hcfirst} with a larger F1 score than the corresponding F1 score threshold. Each subplot shows DRAM modules from a different manufacturer, and each curve represents a different module.} 

\agy{4}{We make three observations from \figref{fig:hcfirst_correlations}.
First, the fraction of spatial features drastically drops when F1 score threshold is increased from 0.6 to 0.7 for \emph{all} modules. 
Second, \emph{no} spatial feature strongly correlates with \gls{hcfirst} when F1 score threshold is chosen as 0.8. 
Third, \emph{only} four modules (S0, S1, S3, and S4) out of \param{15} tested modules have spatial features correlating with \gls{hcfirst} with an F1 score above 0.7 \agy{5}{(not shown in the figure)}.}\footnote{\agy{5}{We empirically choose the threshold of 0.7 to filter out spatial features exhibiting a weak correlation with \gls{hcfirst}
\agy{5}{and provide few stronger features.}}}
{\tabref{tab:correlated_features} shows \agy{5}{the set of spatial features that result in an F1 score above 0.7.} 
Ba, Ro, Sa, and Dist. columns show such spatial features from the bank address, row address, subarray address, and the row's distance to its local sense amplifiers, respectively.} The F1 score column shows the average F1 score for the module across all specified features.

\input{tables/05_correlated_features}

\agy{4}{We make two observations from \tabref{tab:correlated_features}. First,}  
the average F1 score among these features does \emph{not} exceed 0.77 for any tested module. 
\agy{4}{Second, such spatial features mostly come from row and subarray address bits, while \emph{no} bank bit results in an F1 score larger than 0.7. From these two observations, we draw Takeaway~\ref{take:correlations}.} 

\takebox{\agy{4}{Spatial features of DRAM rows \agy{5}{correlate well with their \gls{hcfirst} values in four out of 15 tested DRAM modules.}}\label{take:correlations}}





\subsection{{Repeatability and The Effect of Aging}}
\label{sec:repeatability_and_aging}
\begin{figure}[h!]
    \centering
    \includegraphics[width=0.8\linewidth]{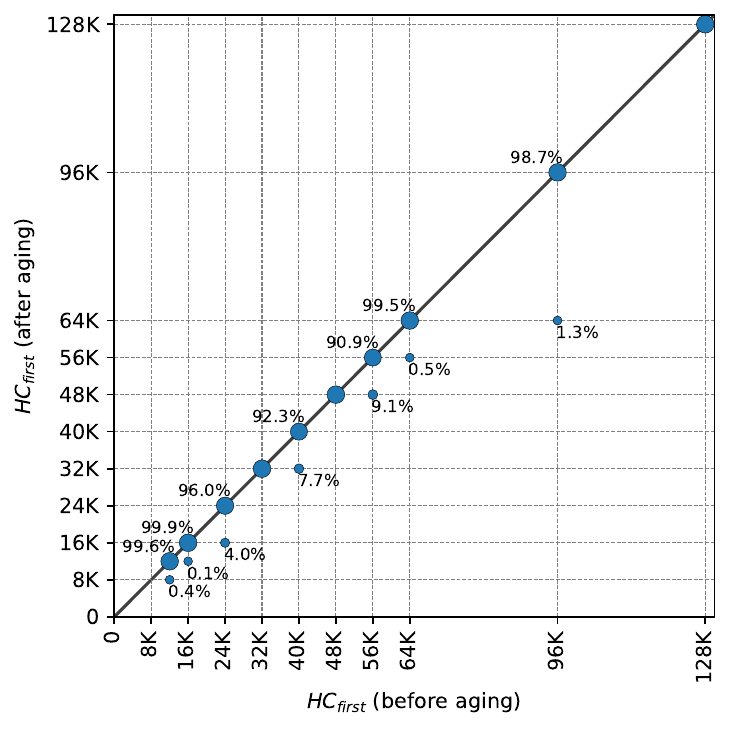}
    \caption{Effect of aging (68 days \agy{4}{using} double-sided RowHammer \agy{4}{test at} \SI{80}{\celsius}) on \gls{hcfirst}}
    \label{fig:aging}
\end{figure}

A DRAM row’s read disturbance vulnerability can change over time. A rigorous aging study requires extensively characterizing many DRAM chips many times over a large timespan. Due to time and space limitations, we leave such studies for future work while presenting a preliminary analysis \agy{4}{on module H3} as our best effort. We repeat our experiments on one of the tested modules after 68 days of keeping the module under double-sided RowHammer tests at \SI{80}{\celsius}. 
\agy{4}{\figref{fig:aging} demonstrates the effect of aging on \gls{hcfirst} in a scatter plot. The x-axis and the y-axis show \gls{hcfirst} values before and after aging, respectively. The size of each marker and annotated text near each data point represent the population of DRAM rows at the given before- and after-aging \gls{hcfirst} values, normalized to the total population of rows at the \gls{hcfirst} before aging, i.e., the population at each x-tick sums up to 1.0. The straight black line marks $y=x$ points, where \gls{hcfirst} does \emph{not} change after aging.}  
We make \obssref{obsv:nonzero_aging} and \ref{obsv:weaker_ages_more} from \figref{fig:aging}. 

\observation{\agy{4}{A non-zero fraction of DRAM rows exhibit lower \gls{hcfirst} values after aging.}\label{obsv:nonzero_aging}} \agy{4}{For example, 0.4\% of DRAM rows with an \gls{hcfirst} of 12K before aging experience bitflips at a \agy{5}{hammer} count of 8K after aging. Therefore, configuring a read disturbance solution for a threshold of 12K is \emph{not} safe for those 0.4\% of the rows, and thus, the \gls{hcfirst} values need to be updated \om{4}{online} in the field.}
We believe this \om{5}{result} makes a \om{5}{strong case} for periodic \om{5}{online} testing of DRAM chips, \om{5}{as also proposed by prior works}~\cite{lee2017design, khan2016parbor, khan2017detecting, liu2013experimental, khan2014efficacy, qureshi2015avatar, patel2017reaper, patel2021harp}.

\observation{Rows with the smallest \gls{hcfirst} values (weakest rows) get affected by aging, unlike the rows with highest \gls{hcfirst} values (strongest rows).\label{obsv:weaker_ages_more}}
For example, \gls{hcfirst} values vary with aging on the left-hand-side of the figure while the rows that show an \gls{hcfirst} value of 128K exhibit no change in their \gls{hcfirst} value. \agy{4}{This indicates that the worst-case \gls{hcfirst} (the lowest \gls{hcfirst} across \om{5}{\emph{all}} rows) changes with aging, and thus aging can jeopardize the security guarantees of existing solutions that are configured based on static \om{5}{identification of the worst-case \gls{hcfirst}}.}
Therefore, finding the correct \gls{hcfirst} under the effect of aging is a challenge for existing solutions \om{4}{and we call for \om{6}{further} research on this topic}.
\agy{4}{Based on \obssref{obsv:nonzero_aging} and \ref{obsv:weaker_ages_more}, we draw Takeaway~\ref{take:aging}.} 

\takebox{Determining \gls{hcfirst} values statically \om{4}{is} \emph{not} completely safe, and finding the worst-case \gls{hcfirst} is an open research problem and challenge for existing solutions due to the variation in minimum \gls{hcfirst} values as a result of aging.\label{take:aging}}


%% file: tables/05_correlated_features.tex

\begin{table}[h!]
  \centering
  \footnotesize
  \caption{Spatial features that correlate with \gls{hcfirst} resulting in an F1 score > 0.7}
    \begin{tabular}{l|llllr}
        {{\bf Module}} & \textbf{Ba} & \textbf{Ro} & {{\bf Sa}}  & {{\bf Dist.}} & {{\bf F1 Score}}\\
        \hline 
        S0 & & Bits 7 and 8 &  Bit 0 & Bit 7 & 0.77\\     
        S1 & & Bits 7, 8, 10, and 12 &  Bit 0 & & 0.71\\     
        S3 & & Bit 10 &  Bits 1 and 2 & & 0.75\\     
        S4 & & &  Bit 0 & & 0.76\\     
        \hline
    \end{tabular}
    \label{tab:correlated_features}
\end{table}

%% file: sections/06_adaptation.tex
\section{{\X{}: Spatial Variation Aware Read Disturbance Defenses}}
\label{sec:adaptation}

We propose a new mechanism \X{}.
The goal of \X{} is to reduce the performance overheads induced by existing read disturbance solutions. \X{} achieves this goal by leveraging the variation in read disturbance vulnerability {across DRAM rows} (\secref{sec:characterization}) to \om{4}{dynamically} tune the aggressiveness of existing read disturbance solutions.

\subsection{High-Level Overview}
\label{sec:adaptation_overview}
\begin{figure}[h!]
    \centering
    \includegraphics[width=\linewidth]{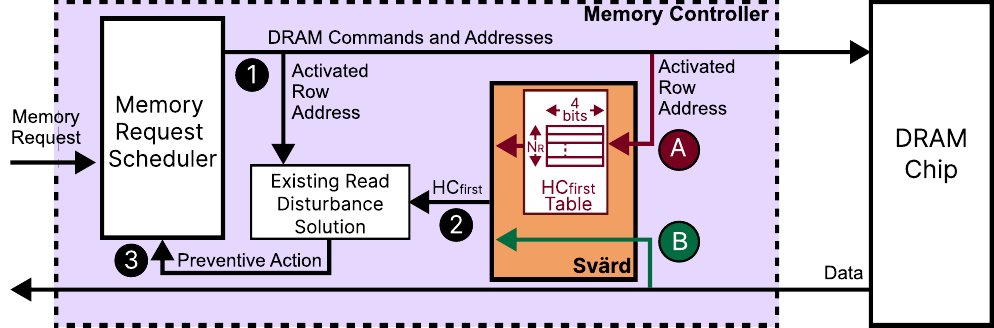}
    \caption{\om{6}{Overview of \X{} MC}-based implementation}
    \label{fig:svard_overview}
\end{figure}

{\figref{fig:svard_overview}\agycomment{4}{ToDo: Revise the figure} shows \X{}'s high-level overview for its memory controller \agy{6}{(MC)}-based implementation. \X{} can similarly be implemented within the DRAM chip (\secref{sec:adaptation_implementation}).}

When a DRAM row is activated \circled{1}, {both \om{4}{an existing} read disturbance solution and \X{} are provided with the activated row address. The existing solution computes \om{4}{a} value (e.g., a random number~\cite{kim2014flipping} or estimated activation count~\cite{qureshi2022hydra, yaglikci2021blockhammer, saxena2022aqua}) to compare against a threshold to decide whether to take a preventive action. Meanwhile \circled{2},} \X{} provides the read disturbance solution with an \gls{hcfirst} value based on the activated row's vulnerability level. Then \circled{3}, the read disturbance solution uses this \gls{hcfirst} value to decide \om{4}{whether or not} to perform a preventive action.
{By providing the \gls{hcfirst} value based on the row's characteristics, \X{} tunes the existing solution's aggressiveness dynamically.} Therefore, the read disturbance solution acts \om{4}{either more or less} aggressively when \om{4}{a row with high or low} vulnerability \om{4}{is} accessed. \om{4}{The} read disturbance solution does \emph{not} perform \om{4}{a} preventive action (e.g., refresh victim rows, throttle accesses to the aggressor row, or relocate the aggressor row's content to a far place from the victim row) if the accessed rows do \emph{not} need the preventive action to avoid bitflips. {\X{} maintains a few \om{5}{bits} (e.g., 4 bits) to specify the \gls{hcfirst} classification of each DRAM row. To do so, \X{} can store and obtain the necessary \om{5}{classification} metadata in various ways, including but not limited to: \coloredcircledletter{burgundy}{A}~implementing an \gls{hcfirst} table within the memory controller \agy{6}{that stores as many entries as the number of rows ($N_{R}$)} and \coloredcircledletter{cadmiumgreen}{B}~fetching the classification data along with the first read from the metadata bits stored in DRAM.} \agy{4}{\om{5}{\X{}'s} classification \om{5}{meta}data storage can be optimized by using Bloom filters, similar to prior work~\cite{liu2012raidr, seshadri2012evicted}.}

\subsection{Implementation Options}
\label{sec:adaptation_implementation}
\X{} can be implemented where the existing read disturbance solution is implemented. \om{5}{We} explain two implementations of \X{} that support \om{4}{read disturbance solutions implemented in} 1)~\om{4}{the} memory controller and 2)~in DRAM \om{4}{chips}. However, there is a large design space for \X{}'s implementation options that we foreshadow and leave for future research.

\head{Memory Controller} Many prior works~\mcBasedRowHammerMitigations{} propose implementing read disturbance solutions in the memory controller where they can observe and enhance all memory requests. To support these solutions, \X{} maintains a data structure that stores the read disturbance profile \om{4}{of DRAM rows} in the memory controller. \X{} observes the row activation ($ACT$) commands that the memory request scheduler issues and uses the activated row address to query the read disturbance profile. In parallel, the read disturbance solution also executes its algorithm (e.g., \om{4}{generates} a random number or \om{4}{increments} the corresponding activation counters). 
\X{} provides the \agy{4}{read disturbance} solution with \om{4}{a more precise} threshold corresponding to the activated row's vulnerability level. \agy{4}{The read disturbance solution uses this more precise threshold to decide whether or not to perform a preventive action.} \X{}'s implementation can follow one of many common practices of storing metadata in computing systems. \X{} can store \om{5}{its} metadata within 1)~a \om{5}{hardware data structure (e.g., a table or Bloom filters)} in the memory controller, 2)~the integrity check bits in the DRAM array~\cite{jedec2020ddr5, patel2019understanding, patel2020beer, patel2021harp, patel2022case, qureshi2021rethinking, fakhrzadehgan2022safeguard}, or 3)~a dedicated memory space in the DRAM array with an optional caching mechanism in the memory controller, \agy{5}{similar to prior works~\cite{qureshi2022hydra, hong2018attache, meza-cal12}}.

\head{DRAM \om{4}{Chip}} \agy{4}{\om{5}{Various} prior works propose to mitigate read disturbance within the DRAM chip}~\inDRAMRowHammerMitigations{}. 
\agy{4}{These read disturbance solutions observe memory access patterns and perform preventive actions within the DRAM chip, \om{5}{transparently} to the rest of the system. To support these read disturbance solutions, \X{} can be implemented within the DRAM chip. Similar to \om{6}{the} memory controller-based implementation, when a DRAM row is activated, \X{} provides the read disturbance solution with a more precise threshold corresponding to the activated row's vulnerability level.}
\X{} can store the necessary metadata within the DRAM array or the activation counters and access when the row is accessed. 

\subsection{Security}
\label{sec:adaptation_security}
{\X{} does \emph{not} \om{4}{affect} the security guarantees of existing read disturbance solutions. 
\agy{4}{Existing read disturbance solutions provide their security guarantees for each DRAM row, conservatively assuming that all DRAM rows are as vulnerable to read disturbance as the most vulnerable (weakest) row.}
As a result, they overprotect the rows that are stronger than the weakest row. With \X{}, these solutions still provide the same security \om{5}{guarantees} for the weakest DRAM rows \om{4}{while at the same time avoiding the overprotection of} the rows that are stronger than the weakest rows without compromising their security guarantees. This is because \X{} tunes the aggressiveness of \om{5}{existing} solutions based on the vulnerability level of \om{4}{each row}.}

\subsection{Hardware Complexity of \om{6}{Storing the} Read Disturbance Vulnerability Profile}

\agy{4}{The hardware complexity of storing the read disturbance vulnerability profile depends on the size of the profile's metadata. The size of this metadata can be reduced by grouping DRAM rows using their spatial features if there is a strong correlation between the spatial features of a DRAM row and the row's \gls{hcfirst}. \secref{sec:spatial_features_correlation_analysis} shows that such correlation exists only for a subset of the tested modules. To make \X{} widely applicable to all DRAM modules, including the ones that do not exhibit such a strong correlation, we evaluate the hardware complexity of storing the read disturbance vulnerability profile for the worst-case, where we cluster rows into several vulnerability bins, and store a bin id separately for each DRAM row.} We evaluate two different implementations of this metadata storage: 1) a table in the memory controller and 2) dedicated bits within the data integrity metadata~\cite{jedec2020ddr5, patel2019understanding, patel2020beer, patel2021harp, patel2022case, qureshi2021rethinking, fakhrzadehgan2022safeguard} in DRAM. In both cases, we store a vulnerability bin identifier per DRAM row. Because the number of bins in each distribution is smaller than 16, we represent each bin with a 4-bit identifier. 
For the area overhead analysis, we assume a DRAM bank size of 64K DRAM rows and a DRAM row size of 8KB. 

First, for the table implementation, we use CACTI~\cite{cacti} and estimate an area cost of \SI{0.056}{\milli\meter\squared} per DRAM bank. When configured for a dual rank system with 16 banks at each rank, the table implementation consumes an overall area overhead of \SI{0.86}{\percent} of the chip area of a high-end Intel Xeon processor with four memory channels~\cite{wikichipcascade}. This table has an access latency of \SI{0.47}{\nano\second}, which can be completely overlapped with the latency of a row activation, e.g., \SI{\approx14}{\nano\second}~\cite{datasheetM393A1K43BB1}.  

Second, storing this metadata as part of the data integrity bits within DRAM requires dedicating four additional bits for an 8KB-large DRAM row. Therefore, it increases the DRAM array size by \SI{0.006}{\percent} with a conservative estimate where each row's width is extended to store four more bits. In this implementation, because the metadata is implemented as part of the data integrity bits, \X{} reads a data word and the metadata in parallel. Therefore, it does \emph{not} increase the memory access latency. \agy{5}{Since the metadata is stored in the DRAM array, the existing read disturbance solution needs to prevent read disturbance bitflips in the metadata. To do so, the read disturbance solution performs its preventive actions (e.g., refreshing a potential victim row) also on the DRAM cells that store the metadata.}

%% file: sections/07_evaluation.tex
\section{Performance Evaluation}
\label{sec:evaluation}

\subsection{Methodology}
\label{sec:evaluation_methodology}

\head{Simulation Environment} {To evaluate the performance impact of \X{}, we conduct {cycle-level simulations} using Ramulator~\cite{ramulatorgithub, Kim2016Ramulator, ramulator2github, luo2023ramulator2}.
Table~\ref{table:configs} shows the simulated system configuration.}

\begin{table}[h]
\centering
\caption{Simulated system configuration}
\label{table:configs}
\resizebox{\linewidth}{!}{
\begin{tabular}{ll}
\hline
\textbf{Processor}                                                   & \begin{tabular}[c]{@{}l@{}} 1 or 8 cores, 3.2GHz clock frequency,\\ 4-wide issue, 128-entry instruction window\end{tabular}  \\ \hline
\textbf{DRAM}                                                        & \begin{tabular}[c]{@{}l@{}}DDR4, 1 channel, 2 rank/channel, 4 bank groups,\\ 4 banks/bank group, 128K rows/bank\end{tabular}  \\ \hline
\begin{tabular}[c]{@{}l@{}}\textbf{Memory Ctrl.}\end{tabular} & \begin{tabular}[c]{@{}l@{}}64-entry read and write requests queues,\\Scheduling policy: FR-FCFS~\cite{rixner2000memory,zuravleff1997controller} \\with a column cap of 16~\cite{mutlu2007stall},\\Address mapping: MOP\cite{kaseridis2011minimalistic}\end{tabular}   \\ \hline
\textbf{Last-Level Cache}& \begin{tabular}[c]{@{}l@{}} 2 MiB per core \end{tabular}  \\ \hline
\end{tabular}}
\end{table}

{In our evaluations, we assume a realistic system with eight cores connected to a memory rank with four bank groups, each containing four banks (16~banks in total). The {memory controller} employs {the} FR-FCFS~\cite{rixner00, zuravleff1997controller} scheduling algorithm with {the} open-{row} policy.}

\head{Comparison Points}
{Our baseline does \emph{not} implement any read disturbance solution and performs memory requests in \om{4}{accordance with} DDR4 specifications~\cite{jedec2017ddr4}. We evaluate \X{} with one of {five} state-of-the-art solutions: \agy{6}{AQUA~\cite{saxena2022aqua}, BlockHammer~\cite{yaglikci2021blockhammer}, Hydra~\cite{qureshi2022hydra}, PARA~\cite{kim2014flipping}, and RRS~\cite{saileshwar2022randomized}.}\textcolor{white}{\ref{fig:mitigation_perf}}


\begin{figure*}[h!]
    \centering
    \includegraphics[width=\linewidth]{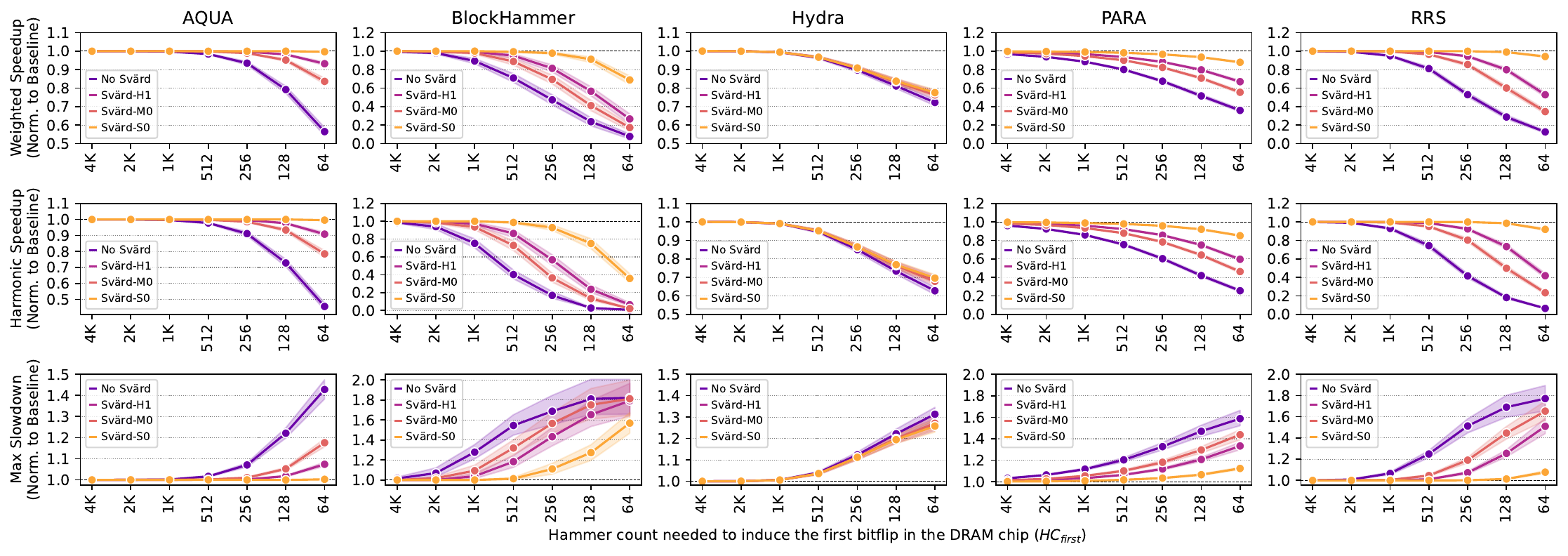}
    \caption{\agy{4}{Performance} overheads of \agy{4}{AQUA~\cite{saxena2022aqua},} {BlockHammer~\cite{yaglikci2021blockhammer},} Hydra~\cite{qureshi2022hydra}, PARA~\cite{kim2014flipping}, and RRS~\cite{saileshwar2022randomized} with and without \X{}}
    \label{fig:mitigation_perf}
\end{figure*}

\head{Workloads}
\copied{HiRA}{We execute {\wlcnt{} 8-core multiprogrammed workload mixes}, randomly chosen from \copied{ABACuS}{{five} benchmark suites: {SPEC CPU2006~\cite{spec2006}}, SPEC CPU2017~\cite{spec2017}, {TPC~\cite{tpcweb}, MediaBench~\cite{fritts2009media}, and YCSB~\cite{ycsb}}}.
We simulate these
workloads until each core executes 200M instructions with a warmup period of 100M, similar to prior work~\cite{kim2020revisiting, yaglikci2021blockhammer, yaglikci2022hira}.}
 
\head{Metrics}
\copied{BlockHammer}{{We evaluate \X{}'s impact on \emph{system throughput} (in terms of weighted speedup~\cite{snavely2000symbiotic, eyerman2008systemlevel, michaud2012demystifying}), \emph{job turnaround time} (in terms of harmonic speedup~\cite{luo2001balancing,eyerman2008systemlevel}), and \emph{fairness} {(in terms of maximum slowdown~\cite{kim2010thread, kim2010atlas,subramanian2014bliss,subramanian2016bliss, subramanian2013mise, mutlu2007stall, subramanian2015application, ebrahimi2012fairness, ebrahimi2011prefetch, das2009application, das2013application, yaglikci2021blockhammer}})}.}

\head{Read Disturbance Vulnerability Profile}
To evaluate \X{}, we use each manufacturer's read disturbance vulnerability profile based on \om{4}{our} real chip characterization results \om{4}{(\secref{sec:characterization})}. Our evaluation spans seven different \agy{4}{worst-case} \gls{hcfirst} values \agy{4}{from 4K down to 64 to evaluate system performance for} modern and future DRAM chips \agy{4}{as technology node scaling over generations exacerbates read disturbance vulnerability. To apply our read disturbance vulnerability profile to future DRAM chips, we scale down all observed \gls{hcfirst} values}
such that the minimum \agy{4}{(worst-case) \gls{hcfirst} value in the read disturbance vulnerability profile becomes} equal to the \gls{hcfirst} value used \agy{4}{for evaluating the read disturbance solution \emph{without} \X{}.} 

\subsection{Performance Analysis}
\label{sec:evaluation_results}
\figref{fig:mitigation_perf} shows how system performance varies when \yct{4}{five} state-of-the-art read disturbance solutions are employed with and without \X{}. Each column of subplots evaluates a different read disturbance solution. We sweep \gls{hcfirst} on the x-axis from 4K down to 64. We show the three performance metrics (weighted speedup, harmonic speedup, and max. slowdown) normalized to the baseline where \agy{4}{\emph{no} read disturbance solution} is implemented. We annotate each configuration of \X{} in the form of \X{}-[DRAM Mfr], \agy{4}{where we choose a representative module from each manufacturer}. Each marker and shade show the average and the minimum-maximum span of performance measurements across \numworkloadmixes{} workload mixes. We make \obssref{obsv:allgood}-\ref{obsv:MfrSgood} from \figref{fig:mitigation_perf}.   

\observation{\X{} consistently improves system performance in terms of all three metrics when used with any of the \yct{4}{five} tested solutions for all \gls{hcfirst} values below 1K.\label{obsv:allgood}} \X{} configurations result in clearly \om{4}{higher} values for weighted and harmonic speedups and lower values for maximum slowdown, \agy{4}{compared to the respective solution \emph{without} \X{}}. 
{For \agy{4}{an \gls{hcfirst} value} of 128 (64), \X{} significantly increases system performance over \yct{4}{AQUA~\cite{saxena2022aqua},} BlockHammer~\cite{yaglikci2021blockhammer}, Hydra~\cite{qureshi2022hydra}, PARA~\cite{kim2014flipping}, and RRS~\cite{saileshwar2022randomized} by
$1.23\times$ ($1.63\times$),
$2.65\times$ ($4.88\times$),
$1.03\times$ ($1.07\times$),
$1.57\times$ ($1.95\times$),
and
$2.76\times$ ($4.80\times$),
respectively, on average across \wlcnt{} evaluated workloads and three read disturbance profiles of modules S0, M0, and H1.}
\X{}'s performance benefits are relatively smaller for Hydra~\cite{qureshi2022hydra} compared to other \agy{4}{evaluated read disturbance solutions}.This is because Hydra's performance overheads are \emph{not} dominated by the preventive refresh operations but the off-chip counter transfer between Hydra's two key components: the counter cache table within the memory controller and the per-row counter table within the DRAM chip. \X{} reduces Hydra's preventive refresh operations, but \emph{not} the off-chip counter transfers. We leave \om{5}{for future work} the Hydra-specific optimizations that \X{} might \agy{4}{enable}.


\observation{\X{} performs the best for the \gls{hcfirst} distribution profile \agy{4}{of S0 among the three evaluated modules.}\label{obsv:MfrSgood}} 
For an \gls{hcfirst} of 64, Svärd reduces \agy{4}{AQUA's~\cite{saxena2022aqua},}
BlockHammer's~\cite{yaglikci2021blockhammer},
Hydra's~\cite{qureshi2022hydra},
PARA's~\cite{kim2014flipping}, and
RRS's~\cite{saileshwar2022randomized} performance overheads of
43.51\%,
92.29\%,
27.75\%,
64.08\%,
87.40\%,
down to
0.32\% / 16.36\% / 6.81\%,
31.15\% / 82.68\% / 73.32\%,
22.44\% / 23.66\% / 22.84\%,
12.05\% / 44.48\% / 33.02\%,
5.83\% / 65.44\% / 47.17\%,
for \agy{4}{modules} S0 / M0 / H1, respectively.
\agy{6}{From \obssref{obsv:allgood} and~\ref{obsv:MfrSgood}, we draw Takeaway~\ref{take:main_perf}.}




\takebox{\X{} effectively reduces the performance degradation that read disturbance solutions inflict on a system.\label{take:main_perf}}

\head{{Adversarial Access Patterns}}
We investigate \X{}'s performance benefits under two adversarial access patterns that exacerbate the performance overheads of Hydra~\cite{qureshi2022hydra} and RRS~\cite{saileshwar2022randomized}. Hydra's adversarial access pattern maximizes the evictions in the counter cache and causes an additional DRAM row activation for each row activation in the steady state. RRS's adversarial access pattern keeps hammering a DRAM row to maximize the number of row swap operations. \figref{fig:adversarial} shows the slowdown caused by Hydra (\figref{fig:adversarial}a) and RRS (\figref{fig:adversarial}b) when these \agy{4}{read disturbance solutions are} used with different \X{} configurations for \agy{4}{an \gls{hcfirst}} of 64. The x-axis shows \X{}'s configurations, and the y-axis shows the measured slowdown (higher is worse), normalized to the slowdown of the \agy{4}{evaluated read disturbance solution \emph{without} \X{} (i.e., No \X{}).}
\agy{5}{We make \obssref{obsv:adv_one} and~\ref{obsv:adv_two}} from \figref{fig:adversarial}.

\begin{figure}[h!]
    \centering
    \includegraphics[width=\linewidth]{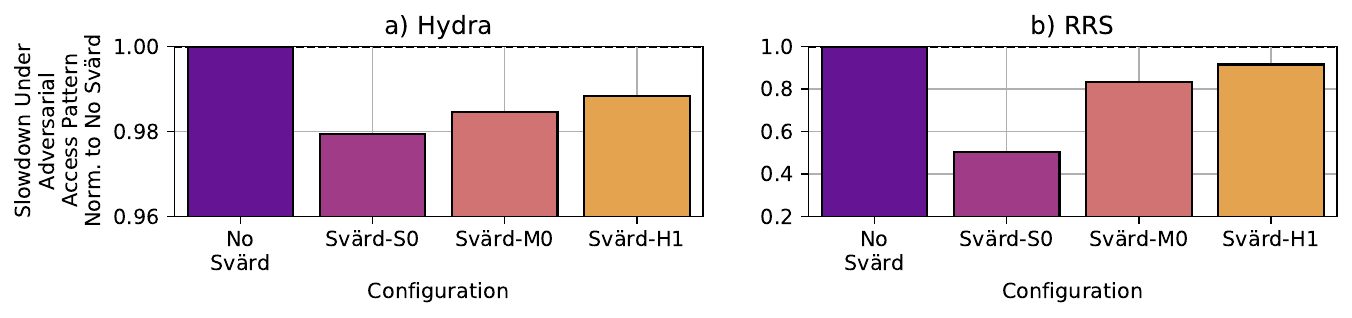}
    \caption{Effect of adversarial access patterns on \X{}'s performance \agy{4}{when used with a) Hydra~\cite{qureshi2022hydra} and b) RRS~\cite{saileshwar2022randomized}}}
    \label{fig:adversarial}
\end{figure}

\observation{\X{} reduces both Hydra's and RRS's performance overheads under adversarial access patterns as all bars except \emph{No \X{}} are below 1.0 for both Hydra and RRS.\label{obsv:adv_one}}
For example, Hydra's and RRS's performance overheads with no \X{} are 73.1\% and 95.6\%, respectively (not shown in the Figure), while \X{} reduces these overheads down to 71.6\% and 48.2\%, respectively, with Mfr. S's profile. 

\observation{Similar to \obsref{obsv:MfrSgood}, \om{4}{\X{} provides the best performance overhead reduction with module S0's profile}
among the three tested profiles.\label{obsv:adv_two}}
The bars for \agy{4}{\X{}-S0} exhibit the lowest slowdown in both \figref{fig:adversarial}a and b. 

From these two \obssref{obsv:adv_one} and~\ref{obsv:adv_two}, we draw Takeaway~\ref{take:adv}.
\takebox{\X{} mitigates the performance overheads of Hydra and RRS under adversarial access patterns.\label{take:adv}}

%% file: sections/09_relatedwork.tex
\section{Related Work}
\label{sec:related_work}

{This is the first work that 
1)~rigorously and experimentally analyzes the spatial variation \om{4}{of} read disturbance in real DRAM chips
and
2)~provide a new mechanism that can dynamically adapt the aggressiveness of existing read disturbance mitigation solutions based on the variation profile of real DRAM chips. 
We highlight our key contributions by contrasting our characterization analysis and proposed mechanism to previous DRAM read disturbance characterization studies and solutions.} 

\head{Characterization Studies}
{Six major works extensively characterize the read disturbance in real DRAM chips~\cite{kim2014flipping,kim2020revisiting,orosa2021deeper, yaglikci2022understanding, olgun2023hbm, luo2023rowpress}.\footnote{{Other prior works~\cite{park2014active, park2016experiments, park2016statistical} present preliminary experimental data from a \om{4}{small number of} (three to five) DDR3 DRAM chips. \om{4}{These works use a} 
small sample set of DRAM cells, rows, and chips.}} 
First, Kim et al.~\cite{kim2014flipping} 
1)~investigate the vulnerability of 129 commodity DDR3 DRAM modules to various RowHammer attack models;
2)~demonstrate for the first time that RowHammer is a real problem
for commodity DRAM chips; and
3)~characterize RowHammer's sensitivity to refresh rate and activation rate in terms of \gls{ber}, \gls{hcfirst}, and the physical distance between aggressor and victim rows.
Second, Kim et al.~\cite{kim2020revisiting} conduct comprehensive scaling experiments on a wide range of 1580 DDR3, DDR4, and LPDDR4 commodity DRAM chips from different DRAM generations and technology nodes, demonstrating that RowHammer has become an even more serious problem over DRAM generations \om{4}{and existing solutions would be expensive for future technology nodes}. 
Third, Orosa et al.~\cite{orosa2021deeper} rigorously characterize various aspects of the RowHammer vulnerability in real DRAM chips, including the effects of temperature, aggressor row active time, and victim DRAM cell's physical location on the RowHammer
vulnerability.  
{Fourth, Ya{\u{g}}l{\i}k{\c{c}}{\i} et al.~\cite{yaglikci2022understanding} investigate RowHammer's sensitivity to wordline voltage and the effects of reducing wordline voltage on DRAM reliability across 272 DRAM chips where 4K rows in a bank is tested in each chip. Fifth, Olgun et al.~\cite{olgun2023hbm, olgun2023hbmarxiv} test \om{5}{an} HBM2 DRAM chip's RowHammer vulnerability and data retention characteristics on a subset of DRAM rows in a DRAM bank with a focus on the spatial variation of RowHammer vulnerability \om{4}{across rows and HBM channels} and the characteristics of on-die target row refresh mechanisms. Sixth, } 
Luo et al.~\cite{luo2023rowpress} are the first to experimentally demonstrate and analyze RowPress, another read disturbance phenomenon, in 164 real DDR4 DRAM chips. They show that RowPress is different \om{4}{from} RowHammer, since it 
1)~affects a different set of DRAM cells from RowHammer and 
2)~behaves differently from RowHammer as temperature and access pattern change.  
{Unfortunately, these} works only analyze \emph{a subset of the DRAM rows.}
In contrast, our {paper takes a} step further and {contributes \om{4}{in} three aspects. First, \agy{4}{we} analyze} 1)~\emph{all} DRAM rows in a DRAM bank and \agy{4}{2)~a DRAM bank from \emph{all} four DRAM bank groups} in a DRAM rank, which {demonstrates the variation in read disturbance vulnerability across rows and banks more accurately \om{4}{and comprehensively}. Second, \om{4}{we study} the correlation between a row’s spatial features and the row’s \agy{4}{read disturbance vulnerability}, which \agy{4}{requires} testing all rows in a DRAM bank. Third, \om{4}{we propose} a \agy{4}{mechanism that leverages our novel experimental observations} to reduce the performance overheads of existing read disturbance solutions.} 
} 


\head{DRAM Read Disturbance Solutions}
\gf{0}{Many prior works propose hardware-based~\hwBasedRowHammerMitigations{} and software-based~\swBasedRowHammerMitigations{} \om{4}{techniques} to prevent RowHammer bitflips, including various approaches, e.g., 
\agy{4}{1)~refreshing potential victim rows \refreshBasedRowHammerDefenseCitations{};
2)~selectively throttling memory accesses \om{5}{that might cause bitflips} \throttlingBasedRowHammerDefenseCitations{};
3)~physically isolating an aggressor row from sensitive data \isolationBasedRowHammerDefenseCitations{}; and}
4)~enhancing DRAM circuitry to mitigate RowHammer effects~\circuitBasedRowHammerDefenseCitations{}. Each approach has different trade-offs \om{5}{in terms of} performance impact, \om{5}{energy overhead, hardware complexity}, and security guarantees. Our goal in this paper is \emph{not} to propose a new read disturbance solution, but rather to provide a \emph{new method} to make the existing solutions more efficient. As we show in \om{5}{\secref{sec:adaptation},} on \agy{4}{\param{five} evaluated} state-of-the-art solutions, \X{} can \emph{significantly} reduce the overheads of read disturbance solutions.}

\head{{\om{5}{RowHammer} Attacks}}
{Many previously proposed RowHammer attacks~\exploitingRowHammerAllCitations{} \om{4}{identify} the rows at the desired vulnerability level (e.g., templating) to increase \agy{4}{the attack's} success probability. This step fundamentally differs from our characterization because it is enough for an attacker to identify a few rows at the desired vulnerability level. In contrast, as a defense-oriented work, we identify the vulnerability level of \om{4}{\emph{all}} DRAM rows \agy{4}{in a DRAM bank} to configure read disturbance solutions correctly without compromising their security guarantees. Therefore, this paper presents a significantly more \om{4}{detailed chip-level} characterization compared to \om{4}{related attack} works.}
    

%% file: sections/10_conclusion.tex
\section{Conclusion}
\label{sec:conclusion}
 
{This paper tackles the shortcomings of existing RowHammer solutions}
{by leveraging the spatial variation of read disturbance vulnerability across different memory locations within a memory module.
To do so, we
1)~present the first rigorous real DRAM chip characterization study of the spatial variation of read disturbance and
2)~propose \X{}, a new mechanism that dynamically adapts the aggressiveness of existing solutions to the read disturbance vulnerability of potential victim rows. 
Our experimental characterization on \numchips{} real DDR4 DRAM chips, {spanning \param{10} different \agy{7}{chip designs}}, demonstrates a large variation in \agy{4}{read disturbance} vulnerability across different memory locations within a module. 
By learning and leveraging this large spatial variation in read disturbance vulnerability across DRAM rows, \X{} reduces the performance overheads of state-of-the-art \agy{4}{DRAM read disturbance} solutions, leading to large system performance benefits. We hope and expect that the understanding we develop via our rigorous experimental characterization and \om{4}{the resulting} \X{} \om{4}{technique} will inspire DRAM manufacturers and system designers to efficiently and scalably \om{4}{enable} robust (i.e., reliable, secure, and safe) operation \om{4}{as} DRAM technology node scaling exacerbates read disturbance.}

%% file: sections/appendix_a_dimms.tex
\section{Tested DRAM Modules}
\label{sec:appendix_tested_dram_modules}
\newcommand*{\myalign}[2]{\multicolumn{1}{#1}{#2}}
\newcommand{\dimmid}[2]{\begin{tabular}[l]{@{}l@{}}#1~\cite{datasheet#1}\\#2~\cite{datasheet#2}\end{tabular}}

\agy{4}{Table~\ref{tab:detailed_info} shows the characteristics of the DDR4 DRAM modules we test and analyze. We provide {the} module and chip identifiers, access frequency (Freq.), manufacturing date (Mfr. Date), {chip} density (Chip Den.), die revision {(Die Rev.)}, chip organization {(Chip Org.), and the number of rows per DRAM bank} of tested DRAM modules. We report the manufacturing date of these modules in the form of $week-year$. For each DRAM module, Table~\ref{tab:detailed_info} shows the minimum (Min.), average (Avg.), and maximum (Max.) \gls{hcfirst} values across all tested rows.}
\renewcommand{\arraystretch}{1.2}
\begin{table*}[ht]
\footnotesize
\centering
\caption{Characteristics of the tested DDR4 DRAM modules.}
\label{tab:detailed_info}
\begin{tabular}{|l|l|l||c|cccc|c|ccc|}
\hline
\textbf{Module} & \textbf{Module} & \textbf{Module Identifier} & \textbf{Freq} & \textbf{Mfr. Date} & \textbf{Chip} & \textbf{Die} & \textbf{Chip} & \textbf{Num. of Rows} & \multicolumn{3}{c|}{\gls{hcfirst}}\\ 
\textbf{Label}&\textbf{Vendor}&\textbf{Chip Identifier}&\textbf{(MT/s)}&\textbf{ww-yy}&\textbf{Den.}&\textbf{Rev.}&\textbf{Org.}&\textbf{per Bank}&\textbf{Min.}&\textbf{Avg.}&\textbf{Max.}\\
\hline
\hline
H0 & & \dimmid{HMAA4GU6AJR8N-XN}{H5ANAG8NAJR-XN} & 3200 & 51-20 & 16Gb & A & $\times8$ & 128K & 16K & 46.2K & 96K\\
\cline{1-1}\cline{3-12}
H1 & & {\dimmid{HMAA4GU7CJR8N-XN}{H5ANAG8NCJR-XN}} & 3200 & 51-20& 16Gb & C & $\times8$ & 128K & 12K & 54.0K & 128K\\
\cline{1-1}\cline{3-12}
H2 & SK Hynix & {\dimmid{HMAA4GU7CJR8N-XN}{H5ANAG8NCJR-XN}} & 3200 & 36-21 & 16Gb & C & $\times8$ & 128K & 12K & 55.4K & 128K\\
\cline{1-1}\cline{3-12}
H3 & & {\dimmid{HMAA4GU7CJR8N-XN}{H5ANAG8NCJR-XN}} & 3200 & 36-21 & 16Gb & C & $\times8$ & 128K & 12K & 57.8K & 128K\\
\cline{1-1}\cline{3-12}
H4 & & \dimmid{KSM32RD8/16HDR}{H5AN8G8NDJR-XNC} & 3200 & 48-20 & 8Gb & D & $\times8$ & 64K & 16K & 38.1K & 96K \\
\hline
\hline
M0 & & \dimmid{MTA4ATF1G64HZ-3G2E1}{MT40A1G16KD-062E} & 3200 & 46-20 & 16Gb & E & $\times16$ & 128K & 8K & 24.5K & 40K\\
\cline{1-1}\cline{3-12}
M1 &  & \dimmid{MTA18ASF2G72PZ-2G3B1QK}{MT40A2G4WE-083E:B} & 2400 & N/A & 8Gb & B & $\times4$ & 128K & 40K & 64.5K & 96K\\
\cline{1-1}\cline{3-12}
M2 & Micron & \dimmid{MTA36ASF8G72PZ-2G9E1TI}{MT40A4G4JC-062E:E} & 2933 &  14-20 & 16Gb & E & $\times4$ & 128K & 8K & 28.6K & 48K \\
\cline{1-1}\cline{3-12}
M3 & & \dimmid{MTA18ASF2G72PZ-2G3B1QK}{MT40A2G4WE-083E:B} & 2400 & 36-21 & 8Gb & B & $\times4$ & 128K & 56K & 90.0K & 128K\\
\cline{1-1}\cline{3-12}
M4 & & \dimmid{MTA4ATF1G64HZ-3G2B2}{MT40A1G16RC-062E:B} & 3200 & 26-21 & 16Gb & B & $\times16$ & 128K & 12K & 42.2K & 96K\\
\hline 
\hline
S0 & & {\dimmid{M393A1K43BB1-CTD}{K4A8G085WB-BCTD}} & {2666} & {52-20} & {8Gb} & {B} & {$\times8$} & 64K & 32K & 57.0K & 128K \\
\cline{1-1}\cline{3-12}
S1 & & {\dimmid{M393A1K43BB1-CTD}{K4A8G085WB-BCTD}} & 2666 & 52-20 & 8Gb & B & $\times8$ & 64K & 24K & 59.8K & 128K\\
\cline{1-1}\cline{3-12}
S2 & Samsung & {\dimmid{M393A1K43BB1-CTD}{K4A8G085WB-BCTD}} & 2666 & 10-21 & 8Gb & B & $\times8$ & 64K & 12K & 42.7K & 96K\\
\cline{1-1}\cline{3-12}
S3 & & \dimmid{F4-2400C17S-8GNT}{K4A4G085WF-BCTD} & 2400 & 04-21 & 4Gb & F & $\times8$ & 32K & 16K & 59.2K & 128K\\
\cline{1-1}\cline{3-12}
S4 & & \dimmid{M393A2K40CB2-CTD}{K4A8G045WC-BCTD} & 2666 & 35-21 & 8Gb & C & $\times4$ & 128K & 12K & 55.4K & 128K\\
\hline 
\end{tabular}

\end{table*}